\documentclass[lettersize,journal]{IEEEtran}
\usepackage{amsmath,amsfonts}
\usepackage{algorithmic}
\usepackage{algorithm}
\usepackage{array}
\usepackage[caption=false,font=normalsize,labelfont=sf,textfont=sf]{subfig}
\usepackage{textcomp}
\usepackage{stfloats}
\usepackage{url}
\usepackage{verbatim}
\usepackage{graphicx}
\usepackage{cite}
\usepackage{booktabs}
\usepackage{multicol}
\usepackage[export]{adjustbox}
\usepackage{xcolor}
\newcolumntype{P}[1]{>{\centering\arraybackslash}p{#1}}

\usepackage{helvet} 

\usepackage{tikz}
\usetikzlibrary{shapes,arrows,positioning,calc} 

\usepackage{afterpage}

\tikzset{
  block/.style = {rectangle, draw, fill=white, text centered, rounded corners, minimum height=2em, minimum width=1em}, 
  block2/.style = {rectangle, draw, fill=white, text centered, rounded corners, minimum height=2em, minimum width=1em},
  input/.style = {coordinate}, 
  output/.style = {coordinate},
  sum/.style = {draw, circle, inner sep=0pt, minimum size=0.5cm}, 
  sum2/.style = {draw, circle, inner sep=0pt, minimum size=0.5cm},
  arrow/.style = {->, thick, >=latex} 
}

\usepackage{bm}

\usepackage{hyperref}

\begin{document}

\bstctlcite{IEEEexample:BSTcontrol} 

\title{An Acoustic Communication Model in Plants}

\author{Fatih~Merdan,
       Ozgur B.~Akan,~\IEEEmembership{Fellow,~IEEE}

\thanks{The authors are with the Center for neXt-generation Communications (CXC), Department of Electrical and Electronics Engineering, Koç University, Istanbul 34450, Türkiye (e-mail: \{fmerdan25, akan\}@ku.edu.tr).}
\thanks{Ozgur B. Akan is also with the Internet of Everything (IoE) Group,
Electrical Engineering Division, Department of Engineering, University of
Cambridge, Cambridge, CB3 0FA, UK (e-mail: oba21@cam.ac.uk).}}

\markboth{}%
{}

\maketitle

\begin{abstract}
Molecular communication (MC) studies biological signals that are found in nature. Most MC literature focuses on particle properties, even though many natural phenomena exhibit wave-like behavior. One such signal is sound waves. Understanding how sound waves are used in nature can help us better utilize this signal in our interactions with our environment. To take a step in this direction, in this paper, we examine how plants process incoming sound waves and take informed actions. Indeed, plants respond to sound, yet no quantitative communication-theoretic model currently explains this behavior. This study develops the first end-to-end acoustic communication framework for plants. The model is formed following the biological steps of the incoming signal, and a mathematical description is constructed at each step following basic biological models. The resulting end-to-end communication-theoretic model is analyzed using MATLAB. Simulations show that a \(200\) \(Hz\), \(20\) \( \mu Pa\) stimulus elevates cytosolic \textbf{\(Ca^{2+}\)} from \(150\) \(nM\) to \textbf{\(230 \pm 10\) \(nM\)} within \(50\) seconds which can cause root bending in plants in the long run. This work establishes quantitative phytoacoustics, enabling bio-inspired acoustic connections for precision agriculture and plant signaling research.

\end{abstract}


\section{Introduction}
\IEEEPARstart{P}{lants} are constantly sensing their environments to make the best use of their available resources \cite{Information_and_Communication_Theoretical_Foundations_of_the_Internet_of_Plants_Principles_Challenges_and_Future_Directions}. To achieve this, they must gather all kinds of information they can from different stimuli like light, temperature, moisture, stiffness of the soil, pH, touch, and \textit{sound waves} which is the focus of this article. Phytoacoustics is the term coined for the study of how plants perceive, produce, and respond to sound waves. Even though the study of sound waves in plant kingdom can be traced back almost to 60 years ago \cite{The_conduction_of_sap_Detection_of_vibrations_produced_by_sap_cavitation_in_Ricinus_xylem}, this topic is gaining more attention these days as the experimental studies conducted yield promising results \cite{Enhancement_of_the_differentiation_of_protocorm-like_bodies_of_Dendrobium_officinale_to_shoots_by_ultrasound_treatment}, \cite{Exposure_to_Sound_Vibrations_Lead_to_Transcriptomic_Proteomic_and_Hormonal_Changes_in_Arabidopsis}, \cite{Flowers_respond_to_pollinator_sound_within_minutes_by_increasing_nectar_sugar_concentration}, \cite{Effects_of_sound-wave_stimulation_on_the_secondary_structure_of_plasma_membrane_protein_of_tobacco_cells}, \cite{Biological_effect_of_sound_field_stimulation_on_paddy_rice_seeds}. However, several key questions remain open—for example, how plants sense sound waves without a dedicated hearing organ, and what internal decision processes govern their reactions to acoustic stimuli. Gaining deeper insights into these mechanisms is essential for integrating plant sensory behavior into Internet of Everything (IoE) based monitoring systems and advancing next-generation precision agriculture technologies \cite{Sustainable_and_Precision_Agriculture_with_the_Internet_of_Everything_IoE}, \cite{Internet_of_Harvester_Nano_Things}.

\begin{figure*}[!t]
  \centering
  \includegraphics[width=0.7\linewidth]{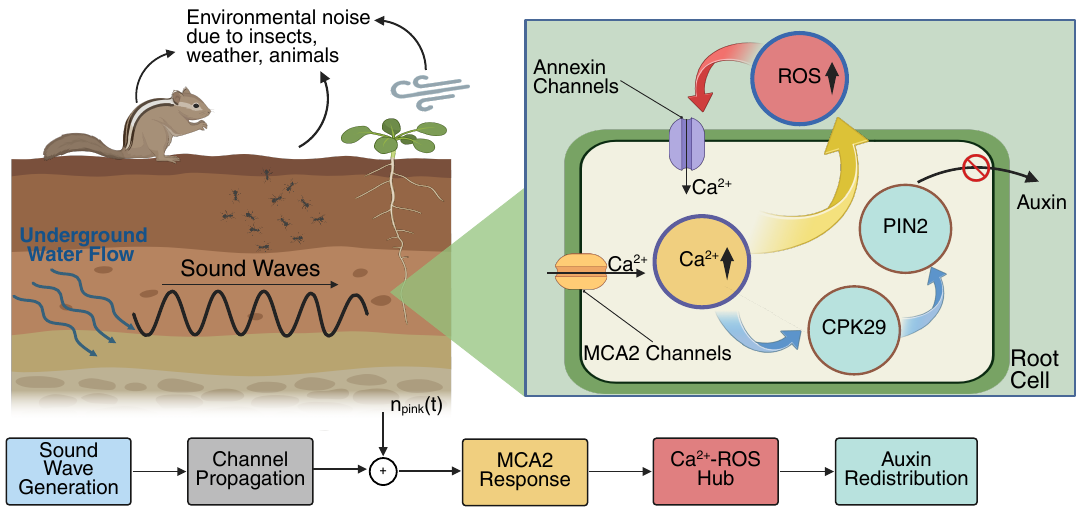}
  \caption{The proposed end-to-end communication-theoretic framework for phytoacoustics. Sound waves activate MCA2 channels, leading to calcium influx with \(Ca^{2+}\)-ROS Hub activity, triggering CPK29 signaling, and modulation of PIN2-mediated polar auxin transport \cite{BioRender_Merdan_2026}.}
  \label{Full_Figure}
\end{figure*}

The objective of this paper is to establish the first communication theoretical model for phytoacoustics in the literature. To the best of our knowledge, there is no communication theoretical approach explaining sound wave transmission and sensing in plants. In this paper, only the sensing of sound stimulus by plants is considered since even though there have been observed some patterns in sounds emitted by plants under stress \cite{Sounds_emitted_by_plants_under_stress_are_airborne_and_informative}, it is still not clear if plants can produce purposeful sound waves. However, it is now well established that plants sense and respond to sound waves \cite{Plant_gene_responses_to_frequency-specific_sound_signals}, \cite{Drought_tolerance_induced_by_sound_in_Arabidopsis_plants}, \cite{Exposure_to_Sound_Vibrations_Lead_to_Transcriptomic_Proteomic_and_Hormonal_Changes_in_Arabidopsis}. Most of the research conducted on plants use the model plant Arabidopsis Thaliana because it has a short life cycle and easy to experiment on. Furthermore, it is the first plant whose genome is identified, and many mutation lines are available \cite{Ion_Channels_in_Plants}, \cite{Analysis_of_the_Genome_Sequence_of_the_Flowering_Plant_Arabidopsis_thaliana}. This paper presents a communication theoretical model that focuses on the roots of the model plant Arabidopsis Thaliana under the sound waves created by water flow through the soil. The main structure of the proposed framework is shown in \autoref{Full_Figure}.

The rest of this paper is organized as follows. In Section II, the acoustic pathway in plants is discussed. In Section III, sound generation and transmission are explained. Sound perception by plants is explained in Section IV. In Section V, the mathematical framework of sound perception by plants is described. In Section VI, the simulation results are given and discussed. In Section VII, communication-theoretic analysis is conducted. The paper is finalized with the concluding remarks in Section VIII.

\section{Acoustic Pathway in Plants}

A sound wave is basically an oscillation of pressure levels. In essence, it can be considered as a mechanical stimuli. Plants sense mechanical stimuli, like touch and wind, through mechanosensing receptors on the plasma membranes of their cells. Mechanoperception mostly occurs by an increase in the cytosolic \(Ca^{2+}\) levels, since these mechanosensing receptors trigger a calcium flux through the cytosol. After sound stimuli, a similar \(Ca^{2+}\) behaviour also occurs in plants \cite{Root_phonotropism_Early_signalling_events_following_sound_perception_in_Arabidopsis_roots}. Therefore, we assumed that hearing in plants is facilitated by mechanically sensible \(Ca^{2+}\) channels.

Calcium channels are membrane proteins that regulate the flow of \(Ca^{2+}\) ions across cellular membranes, enabling signaling and physiological responses in cells. \(Ca^{2+}\) channels can be divided into two; voltage-dependent
\(Ca^{2+}\) permeable channels (VDCCs) and voltage independent \(Ca^{2+}\) permeable channels (VICCs). VDCCs are divided into two groups: depolarization activated \(Ca^{2+}\) permeable channels (DACCs) and hyperpolarization activated \(Ca^{2+}\) permeable channels (HACCs) \cite{Ca2+_channels_and_Ca2+_signals_involved_in_abiotic_stress_responses_in_plant_cells_recent_advances}, \cite{Mechanisms_of_cytosolic_calcium_elevation_in_plants_the_role_of_ion_channels_calcium_extrusion_systems_and_NADPH_oxidase-mediated_ROS-Hub}, \cite{Ways_of_Ion_Channel_Gating_in_Plant_Cells}. Mechanoreceptors are considered in the VICCs group.  ‘Mechanosensitive channels of small (MscS) conductance’-like channels, ‘mid1-complementing activity’ channels (MCAs), Piezo channels, and hyperosmolality-induced \(Ca^{2+}\) channels (OSCA) are the gene families that encode these channels \cite{Deciphering_the_role_of_mechanosensitive_channels_in_plant_root_biology_perception_signaling_and_adaptive_responses}. Today, it is still not exactly clear which of these channels contribute to the perception of sound in plants and to what extent. However, most of them have an established place in plant metabolism. To determine a possible channel, we inspect the directional growth of a plant in response to a mechanical stimulus. Such mechanisms are studied under the term thigmotropism. It is found that MCA channels are important in this context \cite{Determination_of_Structural_Regions_Important_for_Ca2_Uptake_Activity_in_Arabidopsis_MCA1_and_MCA2_Expressed_in_Yeast}, \cite{Transmembrane_Topologies_of_Ca2_permeable_Mechanosensitive_Channels_MCA1_and_MCA2_in_Arabidopsis_thaliana}, \cite{Mix_and_match_Patchwork_domain_evolution_of_the_land_plant-specific_Ca2+-permeable_mechanosensitive_channel_MCA}, \cite{MCA1_and_MCA2_That_Mediate_Ca2_Uptake_Have_Distinct_and_Overlapping_Roles_in_Arabidopsis}. In fact, there are two types of MCA channels: MCA1 and MCA2 who share 72.7\% amino acid sequence identity and several common structural features. MCA1 senses soil hardness and influences root bending to avoid hard soil to facilitate thigmotropism. MCA2 is also important for thigmotropism; however, its exact role relative to MCA1 in plant metabolism remains unclear. Therefore, we propose that MCA2 channels are responsible for sound perception in plant roots to fill this gap and assume it to be the main sensory unit for sound waves in this paper.

The main growth hormone in plants, which is effective in gravitropism and thigmotropism, is auxin \cite{All_Roads_Lead_to_Auxin_Post-translational_Regulation_of_Auxin_Transportby_Multiple_Hormonal_Pathways}, \cite{Rapid_Auxin-Mediated_Cell_Expansion}, \cite{Auxin_steers_root_cell_expansion_via_apoplastic_pH_regulation_in_Arabidopsis_thaliana}. Conversely, abicidic acid is primarily responsible for mediating the response to moisture stimuli \cite{Hydrotropism_how_roots_search_for_water}, \cite{Hydrotropism_mechanisms_and_their_interplay_with_gravitropism}. Considering sound waves as mechanical stimuli, the main hormone of interest in this paper is auxin.


The block diagram of the communication system is given in \autoref{Full_Figure}. Flow of water creates a sound wave which stands for the transmitter of the communication system. Then, this sound wave travels underground and reaches the root cells of the plant where it triggers MCA2 channels, which leads to a \(Ca^{2+}\) influx. \(Ca^{2+}\) influx increases the apoplastic reactive oxygen species (ROS) concentrations of the root cells. Especially high apoplastic \(H_2O_2\) concentrations activates another type of \(Ca^{2+}\) channel which increases cytosolic \(Ca^{2+}\) levels more. This positive feedback mechanism is referred to as the \(Ca^{2+}\) - ROS Hub, which eventually results in the redistribution of auxin in root cells.

\section{Sound Generation and Transmission}

\begin{figure*}[!t]
    \centering
    \includegraphics[width=\textwidth]{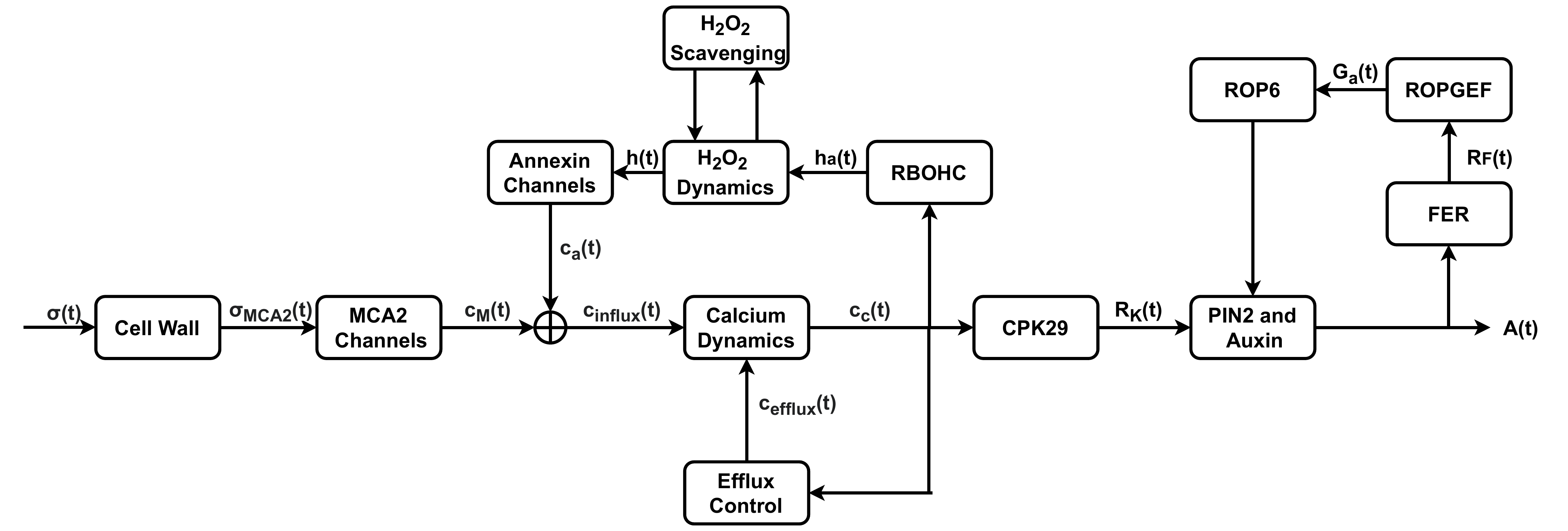}
    \caption{Block diagram showing the proposed model for the sound perception by plants.}
    \label{Plant_hearing_block_model_detailed}
\end{figure*}

Propagation of sound waves under porous media, e.g., soil follows Biot's Theory \cite{Theory_of_Propagation_of_Elastic_Waves_in_a_Fluid_Saturated_Porous_Solid_I_Low_Frequency_Range}, \cite{Theory_of_Propagation_of_Elastic_Waves_in_a_Fluid_Saturated_Porous_Solid_II_Higher_Frequency_Range}. Considering soil as a communication channel, Kelvin–Voigt Model, which takes sound attenuation into account as well as both the viscous and elastic properties of the soil, has been used in this paper \cite{Wireless_Underground_Sensor_Communication_Using_Acoustic_Technology}, \cite{Geotechnical_Earthquake_Engineering}. Therefore, underground sound waves can be expressed as a function of time and space as

\begin{equation}
\label{Sound_underground}
s(x,t) \bm{=} A_s e^{k_2 x} e^{i(\omega t-k_1 x)},
\end{equation}

\noindent with,

\begin{equation}
\label{Sound_underground_k1}
k_1^2 \bm{=} \frac{\rho \omega^2}{2G(1+4\zeta^2)} \left( \sqrt{1+4\zeta^2} + 1 \right),
\end{equation}

\begin{equation}
\label{Sound_underground_k2}
k_2^2 \bm{=} \frac{\rho \omega^2}{2G(1+4\zeta^2)} \left(\sqrt{1+4\zeta^2}-1 \right),
\end{equation}

\begin{equation}
\label{Sound_underground_zeta}
\zeta \bm{=}  \frac{\eta \omega}{2G},
\end{equation}

\noindent where \(t\) is time, \(x\) is position, \(\zeta\) is the damping ratio, \( \omega \bm{=} 2\pi f \) is the angular frequency, \(G\) is shear modulus (\(MPa\)), \(\eta\) is the viscosity of the material (Pas) and \(\rho\) is bulk density (\(gm/cm^3\)). Clay soil texture is assumed to be used and external sources are utilized for related parameters. 


The frequency \(200\) \(Hz\) is associated with water flow, and plant roots are observed to follow the sound waves with this frequency \cite{Towards_understanding_plant_bioacoustics}, \cite{Root_phonotropism_Early_signalling_events_following_sound_perception_in_Arabidopsis_roots}, \cite{Tuned_in_plant_roots_use_sound_to_locate_water}. Therefore, this frequency is used to model sound waves created by water flow. However, in nature, it is not possible for only sound waves with a frequency of \(200\) \(Hz\) to be created during water flow. Changes in the flow rate, interactions with obstacles, and turbulence effects may cause different frequencies to occur as the water flows. To account for this effect, the transmitted sound wave is expressed as

\begin{equation}
\label{Sound_underground_sum}
s(x,t) \bm{=} \sum_{n=1}^{N_1} A_s e^{k_2 x} e^{i(\omega_n t-k_1 x)},
\end{equation}

\noindent where different frequencies can be created at the same time and \(f_n\) in \(\omega_n \bm{=} 2\pi f_n \) is treated as a Gaussian random variable with mean \(200\) \(Hz\) and standard deviation of \(60\) \(Hz\) which is decided by considering the spectrogram of the recorded sound of water flow through a pipe \cite{Tuned_in_plant_roots_use_sound_to_locate_water}. Moreover, a random phase between 0 and \(2\pi\) is added for each sound wave. The amplitude \(A_s\) is also taken as a Gaussian random variable with mean \(20\) \( \mu Pa\) and a standard deviation of \(1\) \( \mu Pa\).

Since Kelvin–Voigt Model already describes the attenuation and phase changes of the propagating sound waves, for expressing the channel response, one should only insert the distance value to (\ref{Sound_underground_sum}).
Therefore, the channel impulse response is modelled as

\begin{equation}
\label{Channel_impulse}
h_c(x) \bm{=} \delta(x-x_m),
\end{equation}

\noindent where \(x_m\) is used to describe the distance between the water and the root cells. Moreover, there could be environmental sounds created by insects, weather, animals, or humans. To account for this effect, pink noise is added to the output of the channel. Pink noise has a power spectrum that decays with 1/\(f\) and this means only low frequency sounds will have a considerable effect. This correlates with sound waves propagating through the soil, as it can be observed in \eqref{Sound_underground} and \eqref{Sound_underground_k2}; the higher the frequency, the higher the decay.

\section{Sound Perception by Plants}

The structures called trichomes are the first to sense the sound wave oscillations, and they work as antennae for sound stimulus \cite{Arabidopsis_Leaf_Trichomes_as_Acoustic_Antennae}, \cite{Acoustic_radiation_force_on_a_long_cylinder_and_potential_sound_transduction_by_tomato_trichomes}, \cite{The_Arabidopsis_Trichome_is_an_Active_Mechanosensory_Switch}. However, they do not exist in the root cells; therefore, when a sound wave reaches the root cells of the plant, the first structure it interacts with is the cell wall. In \cite{Feeling_Stressed_or_Strained_A_Biophysical_Model_for_Cell_Wall_Mechanosensing_in_Plants}, a mathematical model is presented that explains how pressure applied to the cell wall affects the small protein-like structures such as MCA2 channels on the cell membrane. In this paper, this model is utilized. MCA2 channels have been studied previously and it is found that four of these proteins can form a channel structure (tetramer structure) \cite{Structural_Characterization_of_the_Mechanosensitive_Channel_Candidate_MCA2_from_Arabidopsis_thaliana}, \cite{MCAs_in_Arabidopsis_are_Ca2_permeable_mechanosensitive_channels_inherently_sensitive_to_membrane_tensions}. Previous studies provide information about opening probabilities and current flow through MCA2 channel, which enables one to find the increase in the \(Ca^{2+}\) concentration in the cytosol.

Increasing levels of \(Ca^{2+}\) in cytosol triggers NADPH oxidase RBOHC (Respiratory burst oxidase homolog protein C) to activate. \(Ca^{2+}\) binds to the EF-hand domains of the RBOHC which makes the protein undergo a conformational change, which, in the end, activates it \cite{EF_hand_calcium_binding_proteins}, \cite{Mechanisms_of_cytosolic_calcium_elevation_in_plants_the_role_of_ion_channels_calcium_extrusion_systems_and_NADPH_oxidase-mediated_ROS-Hub}. NADPH oxidases are responsible for the production of extracellular ROS (reactive oxygen species) which is another secondary messenger in plant biology. Especially, apoplastic Hydrogen Peroxide (\( H_2O_2 \)) levels increase after the activation of RBOHC. Apoplastic \( H_2O_2 \) binds to annexin channels, which are binding-activated \(Ca^{2+}\) channels on the plasma membrane, and this triggers more \(Ca^{2+}\) to flow into cytosol \cite{Mechanisms_of_cytosolic_calcium_elevation_in_plants_the_role_of_ion_channels_calcium_extrusion_systems_and_NADPH_oxidase-mediated_ROS-Hub}. This positive feedback mechanism makes sure that there is a high \(Ca^{2+}\) concentration inside the cell \cite{Powering_the_plasma_membrane_Ca2-ROS_self-amplifying_loop}, \cite{Calcium_transport_across_plant_membranes_mechanisms_and_functions}.

High cytosolic \(Ca^{2+}\) also activates the CPK29 (calcium-dependent protein kinase 29) by a similar EF-hand \(Ca^{2+}\) binding mechanism. CPK29 is a calcium-dependent protein kinase that functions in the phosphorylation of PIN2 proteins \cite{Calcium-dependent_protein_kinase_29_modulates_PIN-FORMED_polarity_and_Arabidops_is_development_via_its_own_phosphorylation_code}. PIN family of proteins play a critical role in the regulation of auxin transport in plants. 

Auxin levels of the cells of the Arabidopsis Thaliana are regulated by mainly two types of transporters. The entry of auxin into cells is primarily mediated by AUX1 transporters. Since most cells have symmetric distributions of AUX1, polar auxin transport is achieved by regulation of PIN carriers, which are the main efflux transporters of auxin \cite{All_Roads_Lead_to_Auxin_Post-translational_Regulation_of_Auxin_Transportby_Multiple_Hormonal_Pathways}. Therefore, the regulation of PIN2 localization may result in polar auxin transport, which eventually causes one side of the root to grow longer, leading to bending.

There are two main mechanisms in the regulation of PIN proteins: Endocytosis refers to the removal of PINs from the plasma membrane, temporarily reducing auxin transport out of the cell in a specific direction. Recycling of PINs refers to the process by which PIN proteins are continuously cycled between the plasma membrane and internal cellular compartments. It is observed that CPK29 directly binds to PIN2 and an energy transfer occurs between the two \cite{Calcium-dependent_protein_kinase_29_modulates_PIN-FORMED_polarity_and_Arabidops_is_development_via_its_own_phosphorylation_code}. However, it is not established whether this results in the endocytosis of PIN2 proteins, or recycling of PIN2 proteins back to the plasma membrane. In this paper, we assume that CPK29 activity results in the endocytosis of PIN2 proteins. Because, this way, the side of the root that is closer to the water flow will not have much auxin in its root tips, and the other side will grow longer. This results in the bending of the root towards the water.

There is another pathway that influences PIN2 localization, which is triggered by apoplastic auxin \cite{Mathematical_Modelling_of_Auxin_Transport_in_Plant_Tissues_Flux_Meets_Signalling_and_Growth}. Apoplastic auxin triggers the receptor like-kinase FERONIA to be activated by binding to it. Activated FERONIA directly engages with Rop-guanine nucleotide exchange factors (RopGEFs), which promote the exchange of GDP for GTP, transitioning RAC/ROPs from their inactive GDP-bound state to their active GTP-bound state \cite{FERONIA_and_Her_Pals_Functions_and_Mechanisms}. In this paper, ROP6 is considered because it is found that it causes the inhibition of PIN2 internalization which means it blocks endocytosis of PIN2 \cite{A_ROP_GTPase-Dependent_Auxin_Signaling_Pathway_Regulates_the_Subcellular_Distribution_of_PIN2_in_Arabidopsis_Roots}, \cite{Activation_of_ROP6_GTPase_by_Phosphatidylglycerol_in_Arabidopsis}.

\section{Mathematical Modelling of Sound Perception by Plants}

The proposed model for sound perception by plants is given in \autoref{Plant_hearing_block_model_detailed}. As described in \cite{Feeling_Stressed_or_Strained_A_Biophysical_Model_for_Cell_Wall_Mechanosensing_in_Plants}, using the stress on the cell walls, one can find the force applied on MCA2 proteins, $F(t)$. It can be expressed as

\begin{equation}
\label{force_from_stress}
F(t) \bm{=} \frac{\mu_s}{\mu_w} \left( \frac{\tau_w}{\tau_s} \Delta_{\tau_s} \tilde{\sigma}(t) + \langle \tilde{\sigma}(t) \rangle_{\tau_s} \right),
\end{equation}

\noindent with 

\begin{equation}
\label{force_from_stress_extra1}
\langle \tilde{\sigma}(t) \rangle_{\tau_s} = \int_{-\infty}^0 \frac{d\tau}{\tau_s} e^{\frac{\tau}{\tau_s}} \tilde{\sigma}(t+\tau),
\end{equation}

\begin{equation}
\label{force_from_stress_extra2}
\Delta_{\tau_s} \tilde{\sigma}(t) = \tilde{\sigma}(t) - \langle \tilde{\sigma}(t) \rangle_{\tau_s},
\end{equation}

\begin{equation}
\label{force_from_stress_extra3}
\tilde{\sigma}(t) \bm{=} \sigma - Y,
\end{equation}

\noindent where the cell wall is modeled as a viscoelastoplastic material consisting of a spring of elastic modulus \(K_w\) in series with a dashpot of viscosity \(\mu_w\), arranged in parallel with a frictional block that has a yield stress Y, which is the stress beyond which the wall begins to exhibit viscous behavior. \(\tau_w = \mu_w/K_w\) is the relaxation time. The sensor (MCA2) is assumed to be embedded in a visco-elastic medium, arranged in parallel with the wall, and characterized by an elastic modulus \(K_s\), and a viscous coefficient \(\mu_s\). The relaxation time, \(\tau_s = \mu_s/K_s\), defines the point at which the medium begins to display viscous behavior. \(\langle \tilde{\sigma}(t) \rangle_{\tau_s}\) is the slow tend of stress \(\tilde{\sigma}(t)\).

To find the pressure applied on one MCA2 channel \(\sigma_{MCA2}(t)\), the resulting force is divided by the surface area of the MCA2 channels, \(A_M\) which is calculated by assuming MCA2s as cylinders, taking into account the tetramer structure of the channels and the size values from \cite{Structural_Characterization_of_the_Mechanosensitive_Channel_Candidate_MCA2_from_Arabidopsis_thaliana}.


\subsection{MCA2 Channels}

\begin{figure}[!ht]
    \centering
    \includegraphics[width=0.4\textwidth]{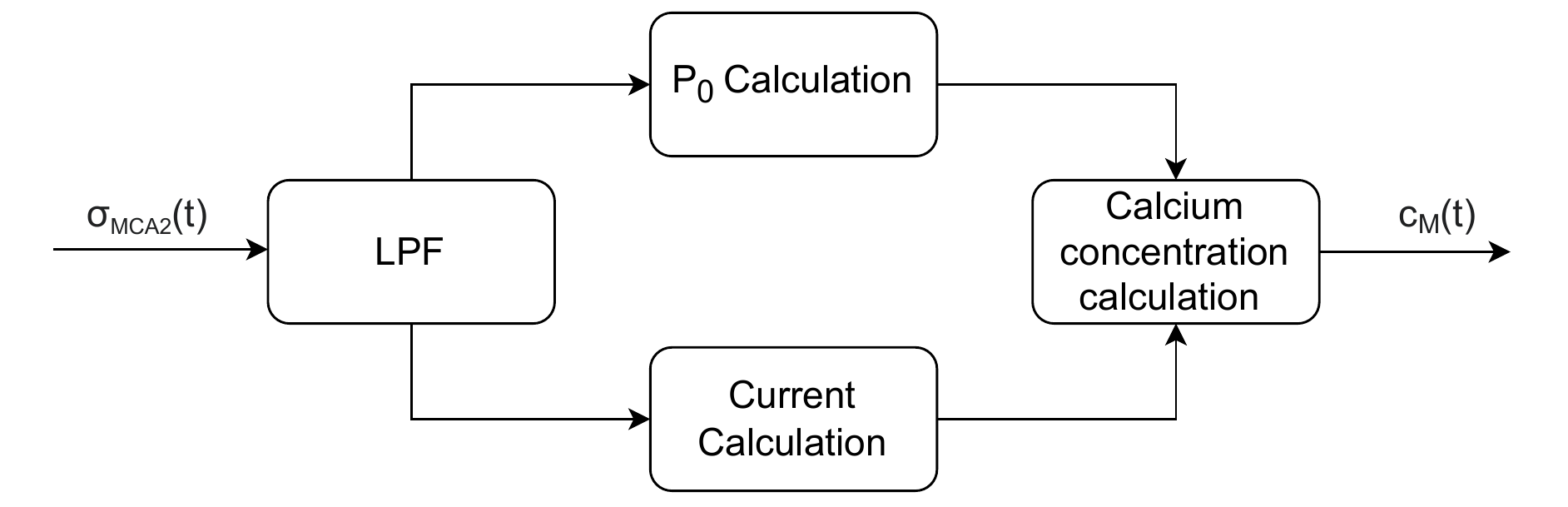}
    \caption{Block diagram modelling MCA2 channels.}
    \label{MCA2_block_model}
\end{figure}

MCA2 block in \autoref{Plant_hearing_block_model_detailed} can be further expanded as shown in \autoref{MCA2_block_model}. MCA2 channels have activation, inactivation, and deactivation phases. This mechanism can be supported by the study in \cite{MCAs_in_Arabidopsis_are_Ca2_permeable_mechanosensitive_channels_inherently_sensitive_to_membrane_tensions}, where current curves have almost periodic spikes and zero values between the spikes indicate the channel being closed for some time. This phenomenon results in higher-frequency stimuli being eliminated. Therefore, the first block in the MCA2 block diagram is a low-pass filter.

Using the data provided in \cite{MCAs_in_Arabidopsis_are_Ca2_permeable_mechanosensitive_channels_inherently_sensitive_to_membrane_tensions}, the probability of the MCA2 channel being in the open state is observed to be exponentially dependent on voltage. Such relationships can be modeled using the Boltzmann distribution, and a reasonable curve is fit to the measurement data shared in \cite{MCAs_in_Arabidopsis_are_Ca2_permeable_mechanosensitive_channels_inherently_sensitive_to_membrane_tensions}. The resulting function describing the voltage dependence of the probability of open state, \(P_0\), is given as


\begin{equation}
\label{Prob_voltage_MCA2}
P_0(V) \bm{=} \frac{1}{1+e^{ \frac{V_{h} - V}{k_V} } },
\end{equation}

\noindent where \(V_{h}\) corresponds to the voltage at which the probability of the MCA2 channel being in the open state is \(50\)\%. Similarly, the pressure dependence of this probability is found and given as

\begin{equation}
\label{Prob_pressure_MCA2}
P_0(\sigma) \bm{=} \frac{1}{1+e^{ \frac{\sigma_{h} - \sigma}{k_\sigma} } },
\end{equation}
 
\noindent where \(\sigma_{h}\) corresponds to the stress level at which the probability of the MCA2 channel being in the open state is \(50\)\%. Using the fact that pressure data is obtained for \(150\) \(mV\), the total open state probability function is described as

\begin{equation}
\label{Prob_MCA2}
P_0^{MCA}(V,\sigma) \bm{=} l \times \frac{1}{1+e^{ \frac{V_{h} - V}{k_V} } } \times \frac{1}{1+e^{ \frac{\sigma_{h} - \sigma}{k_\sigma} } }.
\end{equation}

\(Ca^{2+}\) will flow into the cytosol when MCA2 channels become open because of both the concentration gradients and electric fields. Therefore, the current that passes through the MCA2 channels can be expressed using the Nernst-Planck electrodiffusion equation \cite{Ion_Channels_of_Excitable_Membranes}. Here, the derivatives are simplified into changes in the electric potential and concentration difference of \(Ca^{2+}\) ions, i.e.,

\begin{equation}
\label{Current_MCA2}
I_M \bm{=} -z_{Ca} F D_{Ca} (\frac{\Delta c}{\Delta x_d} + \frac{F z_{Ca} c_c}{RT} \frac{\Delta E}{\Delta x_d} ) \times A_C ,
\end{equation}

\noindent with diffusion coefficient \(D_{Ca}\) is modified so that it depends on the applied pressure, i.e.,

\begin{equation}
\label{Current_MCA2_diffusion}
D_{Ca} \bm{=} \frac{RT}{F} u_{Ca} ( 1+ k_d \sigma_f ),
\end{equation}

\noindent where \( z_{Ca} =2 \) is the valence of ion \(Ca^{2+}\), R is the universal gas constant, T is the absolute temperature, F is the Faraday's constant, \(u_{Ca}\) is the mobility of ions, \(\sigma_f\) is the filtered pressure on one MCA2 channel, \(\Delta c\) is the concentration gradient and \(\Delta E\) describes the potential of the cell membrane, \(\Delta x_d\) is the thickness of the cell membrane and \(A_C\) is the surface area where \(Ca^{2+}\) ions can flow through. In \cite{Ion_Channels_of_Excitable_Membranes}, \(D_{Ca} = 0.79*10^{-5} cm^2/s\) is provided without the dependence of \(\sigma_f\). Normally, \(Ca^{2+}\) concentration in the extracellular space is about 1–10 mM, where inside the cell it is between 100-200 nM in resting conditions \cite{Towards_a_Unified_Theory_of_Calmodulin_Regulation_Calmodulation_of_Voltage-Gated_Calcium_and_Sodium_Channels}, \cite{Towards_the_Physics_of_Calcium_Signalling_in_Plants}. Because of this significant difference, \(\Delta c\) can be considered constant throughout the process. Even though \(\Delta E\) may change and there may occur an electrical signal on the plasma membrane (action potentials or variation potentials), modeling it requires considering all kinds of ionic movements like \(K^+\), \(Mg^{2+}\) \cite{ROS_Calcium_and_Electric_Signals_Key_Mediators_of_Rapid_Systemic_Signaling_in_Plants}. In this, for the sake of simplicity, \(\Delta E\) is also considered as a constant.

Having found the probability of the MCA2 channel being in the open state, and the current it creates, the concentration of the \(Ca^{2+}\) ions that are added to the cytosol can be found using the total probability theorem as

\begin{equation}
\label{Ca2_MCA2}
c_M(t) \bm{=} \frac{I_M t}{ z_{Ca} V } \times P_0 \times n_C,
\end{equation}

\noindent where \(V\) is the cytosolic volume of the root cells and \(n_C\) is the number of MCA2 channels.

\subsection{Calcium Dynamics}

Steady state cytosolic calcium concentration is taken as \(c_{ss} = 150\) \(nM\) in root cells. The main calcium influx sources are the MCA2 and annexin channels. These channels are separately studied in the model. There are many mechanisms that decrease the cytosolic calcium concentration. For example, EF-hands of many calcium protein kinases or other proteins accumulate some \(Ca^{2+}\), excess calcium can diffuse to neighboring cells through the structures called plasmodesmata \cite{Controlling_intercellular_flow_through_mechanosensitive_plasmodesmata_nanopores}, \cite{Orchestrating_Rapid_Long-Distance_Signaling_in_Plants_with_Ca_2+_ROS_and_Electrical_Signals}. Moreover, all organelles can store some amount of calcium; however, the vacuole and endoplasmic reticulum (ER) serve as the largest calcium storage sites \cite{The_signatures_of_organellar_calcium}. In this paper, we assumed the following simplified efflux control block

\begin{equation}
\label{Ca_efflux}
c_{efflux}(t) =
\left\{
\begin{array}{ll}
\scriptstyle Q, & \scriptstyle \text{if } c_c(t) \geq 200 \\[4pt]
\scriptstyle k_{eff_2} (c_c(t) - 155), & \scriptstyle \text{if } 165 \leq c_c(t) < 200
\end{array}
\right.,
\end{equation}

\noindent where \(Q = \text{max}(k_{eff_1} (c_c(t) - 195 ),k_{eff_2} (c_c(t) - 155 ) ) \) and \(c_c\) is the cytosolic \(Ca^{2+}\) concentration in \(nM\), so that under high difference from the steady state value, the decrease of \(Ca^{2+}\) will be higher as observed in nature. Calcium Dynamics block basically applies the following starting from the time instance zero and where \(t'\) represents the previous time instance

\begin{equation}
\label{Ca_dynamics}
c_{c}(t) \bm{=} c_{c}(t') + c_{influx}(t) - c_{efflux}(t).
\end{equation}

\subsection{\( H_2O_2 \text{ Dynamics} \)}

Cytoplasmic calcium binds to the EF-hands of the RBOHC and activates it. Activation of a larger protein by smaller structures can be modeled using the Hill equation. Considering RBOHC has 2 EF-hand structures, and two \(Ca^{2+}\) can bind to it simultaneously, the Hill equation is of the form

\begin{equation}
\label{RBOHC_Hill}
R_C \bm{=} \frac{c_c^2}{ k_C + c_c^2},
\end{equation}

\noindent where \(R_C\) represents the fraction of \(Ca^{2+}\) bound (activated) RBOHCs. The total RBOHC concentration in Arabidopsis Thaliana is calculated using (\ref{ppm_relation}) established in \cite{Protein_Abundance_Biases_the_Amino_Acid_Composition_of_Disordered_Regions_to_Minimize_Non-functional_Interactions} where related ppm values are obtained from online databases.

\begin{equation}
\label{ppm_relation}
C_{total} = \frac{k_{ppm} \, AB}{N_A},
\end{equation}

\noindent where \(C_{total}\) is the total concentration, \(N_A\) is the Avagadro Constant, and AB is the abundance value where 1ppm means \(AB=10^{-6}\). This idea is also used to find the total concentrations of CPK29, FERONIA, PIN2, ROPGEF4, and ROP6 proteins. These values are provided in \autoref{Calculated_Values_for_the_Model}. Using the total RBOHC concentration and \(R_C\), one can find the activated RBOHC concentration, \( C_{ac} \) by multiplying the two.

In this paper, the Michaelis-Menten equation which is a general relation describing how enzymes catalyze reactions is used to model the \( H_2O_2 \) generation by RBOHC \cite{Enzymes_Catalysis_Kinetics_and_Mechanisms}

\begin{equation}
\label{H2O2_production}
\upsilon_H \bm{=} \frac{\upsilon_{Hm} C_{ac}(t)}{m_H + C_{ac}(t)},
\end{equation}

\noindent where \(\upsilon_H\) is the reaction rate (velocity) of the process, \(\upsilon_{Hm}\) is the maximum reaction velocity and \(m_H\) is the Michaelis constant for this reaction. Integrating \(\upsilon_H\) with time gives the concentration of \(H_2O_2\) in apoplast created by RBOHC, \(h_a\).

\(H_2O_2\) scavenging refers to the breakdown of hydrogen peroxide by apoplastic catalases. This enzymatic reaction can also be modeled using the Michaelis-Menten equation as follows

\begin{equation}
\label{H2O2_scavenging}
\upsilon_s \bm{=} \frac{\upsilon_{sm} h(t)}{m_s + h(t)},
\end{equation}

\noindent where \(\upsilon_s\) is the reaction rate of the scavenging, \(\upsilon_{sm}\) is the maximum reaction velocity, \(h(t)\) is the apoplastic \(H_2O_2\) concentration, \(m_s\) is the Michaelis constant for this reaction. Integrating \(\upsilon_s\) with time gives the broken down \(H_2O_2\) concentration, \(h_s\).

Inside the \(H_2O_2\) Dynamics block the following operation is performed and apoplastic \(H_2O_2\) concentration \(h(t)\) is given as the output

\begin{equation}
\label{H2O2_dynamics}
h(t) \bm{=} h(t') + h_a(t) - h_s(t)
\end{equation}

\noindent where \(t'\) represents the previous time instance.

\subsection{Annexin Channels}

Annexin channels are ligand-gated channels activated by apoplastic \(H_2O_2\). Such channels can be modeled by the following relation \cite{Towards_the_Physics_of_Calcium_Signalling_in_Plants}

\begin{equation}
\label{CNGC3_current}
I_a \bm{=} G_{a} \frac{[h(t)]^z}{[h(t)]^z + k_a^z} (E_c - \Delta E),
\end{equation}

\noindent where \(G_{a}\) is the channel conductance of the annexin channel, \(\Delta E\) is the cell membrane voltage, \(k_a^z\) is the ligand concentration producing a half-maximal response, and \(z\) is the Hill coefficient, which theoretically represents the number of binding sites. The Nernst equation, i.e.,

\begin{equation}
\label{Ca2+_voltage}
E_c \bm{=} \frac{RT}{z_{Ca} F} ln\frac{c_{ap}}{c_c(t)},
\end{equation}

\noindent can be used to determine the membrane voltage, \(E_c\), known as the equilibrium or reversal potential, which would keep the apoplastic (\(c_{ap}(t)\)) and cytosolic (\(c_c(t)\)) \(Ca^{2+}\) concentrations in a steady state \cite{Towards_the_Physics_of_Calcium_Signalling_in_Plants}.

Using the current relation of annexin channels, one can calculate the concentration of the \(Ca^{2+}\) ions that are added to the cytosol by

\begin{equation}
\label{Ca2_CNGC3}
c_a(t) \bm{=} \frac{I_a t}{ z_{Ca} V } \times n_a,
\end{equation}

\noindent where \(V\) is the cytosolic volume of the root cells and \(n_a\) is the number of annexin channels.

\subsection{CPK29}

Similar to \(Ca^{2+}\) activation of RBOHC, \(Ca^{2+}\) activation of CPK29 can be modeled using Hill equation taking into account that CPK29 has 4 EF-hand structures

\begin{equation}
\label{CPK29_Hill}
R_K \bm{=} \frac{c_c^4}{ k_K + c_c^4},
\end{equation}

\noindent where \(R_K\) represents the fraction of \(Ca^{2+}\) bound (activated) CPK29s. Multiplying this result with the total CPK29 concentration given in \autoref{Calculated_Values_for_the_Model}, one can find the activated CPK29 concentration, \(K_{ac}\). Then, CPK29 phosphorylates PIN2, a process that can be described using Michaelis-Menten kinetics. However, in this case, the Goldbeter-Koshland function can be applied, as described in \cite{Sniffers_buzzers_toggles_and_blinkers_dynamics_of_regulatory_and_signaling_pathways_in_the_cell}

\begin{equation}
\label{CPK29_PIN_Phosp}
\frac{P_n}{P_T} \bm{=} G(\nu_1,\nu_2,j_1,j_2) ,
\end{equation}

\noindent where

\begin{equation}
\label{Goldbeter-Koshland}
G(\nu_1,\nu_2,j_1,j_2) \bm{=} \frac{2 \nu_1 j_2}{ B + \sqrt{B^2 - 4(\nu_2 - \nu_1) \nu_1 j_2 } },
\end{equation}

\begin{equation}
\label{Goldbeter-Koshland_long}
B \bm{=}  \nu_2 - \nu_1 + j_1 \nu_2 + j_2 \nu_1.
\end{equation}

In this formulation, \(P_n\) and \(P_T\) are the inactive and total PIN2 concentrations respectively. \(\nu_1\) and \(\nu_2\) are related to dephosphorylation and phosphorylation rates respectively. \(j_1\) and \(j_2\) are related to Michaelis–Menten constants for dephosphorylation and phosphorylation, respectively. In the calculation, \(\nu_2\) (labelled as \(\nu_2^P\)) is taken directly proportional to the activated CPK29 concentration, \(K_{ac}\) shown as

\begin{equation}
\label{nu_2_CPK29}
\nu_2^P \bm{=}  8E11 \times K_{ac}.
\end{equation}

Subtracting \(\nu_2^P\) from \(1\) and multiplying it with the total PIN2 concentration given in \autoref{Calculated_Values_for_the_Model}, one can find the activated PIN2 concentration, \(P_a\). The values for \(\nu_1^P\), \(j_1^P\) and \(j_2^P\) are given in \autoref{Fixed_Arbitrarily_Chosen_Values_for_Model_Parameters}.

\subsection{PIN2 and auxin}

In \cite{Mathematical_Modelling_of_Auxin_Transport_in_Plant_Tissues_Flux_Meets_Signalling_and_Growth}, a very detailed mathematical model representing PIN and auxin distributions is presented which is also used in this block. For the purposes of this article, a grid of 11x11 cells is established as shown in \autoref{Grid_new}. The parameter \( \alpha_a \), which describes the auxin biosynthesis rate, is doubled for the middle cells to represent auxin transport from the xylem. The left side of these cells is assumed to be hearing the voice and responding to it, while the other cells do not show any reaction. Response to the sound is modeled for the left-sided cells by a modification in their parameter \(\delta_p\), which describes PIN2 membrane dissociation rate, as follows

\begin{table}[!ht]
  \caption{Parameters Extracted from Literature}
  \centering
  \label{PIN2_aux_ready}

  \renewcommand{\arraystretch}{0.7}

  \begin{tabular}{@{}>{\raggedright\arraybackslash\itshape}p{4.2cm}>{\raggedright\arraybackslash\itshape}p{0.6cm}>{\raggedright\arraybackslash\itshape}p{0.5cm}>{\raggedright\arraybackslash\itshape}p{2.1cm}@{}}
    \toprule

 \textbf{Description}   &  \textbf{Symbol}    & \textbf{Value}  & \textbf{Unit}\\
 \midrule
Bulk Density         &\(\rho\)              & 1.30  & gm/\(cm^3\) \\ 
Viscosity         &\(\eta\)              & 1019  & Pas \\ 
Shear Modulus         & G              & 2.4  & MPa \\ 
Auxin degradation rate         &\(\mu_a\)              & 0.5  & \(min^{-1}\) \\  
auxin-TIR1 dissociation rate   & \(\gamma_{a}\)       & 5    &\(min^{-1}\)\\
auxin-TIR1 binding rate     & \(\beta_{a}\) & 0.5        &  \(\mu M^{-1} min^{-1}\)  \\
Maximum mRNA transcription rate    & \(\alpha_{m}\)   & 0.5   &  \(\mu M min^{-1}\) \\
Ratio of ARF-dependent to \(ARF_2\)- and double ARF-dependent mRNA transcription rates         & \(\phi_m\)             & 0.1     & - \\
ARF-DNA binding threshold     &  \(\theta_f\)            & 1  &  \(\mu M\) \\
\(ARF_2\) binding threshold      &  \(\theta_w\)         & 10 &  \(\mu M\) \\
ARF + Aux/IAA-DNA binding threshold       &   \(\theta_g\)      & 1  & \(\mu M\)  \\
Double ARF-DNA binding threshold       & \(\psi_f\)     & 0.1  &  \(\mu M^2\)  \\
ARF-Aux/IAA-DNA binding threshold        &  \(\psi_g\)   & 0.1 &  \(\mu M^2\) \\
Aux/IAA translation rate     &  \(\alpha_r\)     & 5  & \(min^{-1}\)  \\
Aux/IAA-auxin-TIR1 binding rate    & \(\beta_r\)    & 5  & \(\mu M^{-1} min^{-1}\) \\
Aux/IAA-auxin-TIR1 dissociation rate    & \(\gamma_r\)   & 5   &   \(min^{-1}\) \\
ARF-Aux/IAA binding rate      & \(\beta_g\)  & 0.5  &  \(\mu M^{-1} min^{-1}\) \\
ARF-Aux/IAA dissociation rate     & \(\gamma_g\)  & 5  & \(min^{-1}\)  \\
PIN2 translation rate     &   \(\alpha_p\)    & 5     &  \(min^{-1}\) \\
PIN2-auxin-TIR1 binding rate      & \(\beta_p\)  & 100      &  \(\mu M^{-1} min^{-1}\)     \\
PIN-auxin-TIR1 dissociation rate    & \(\gamma_p\)  & 5       &  \(min^{-1}\)     \\
PIN decay rate        &    \(\mu_p\)          & 5     &   \(min^{-1}\)     \\
ARF dimerisation rate     &  \(\beta_f\)  & 0.5     &  \(\mu M^{-1} min^{-1}\)     \\
\(ARF_2\) splitting rate      & \(\gamma_f\)   & 5       &  \(min^{-1}\)     \\
Aux/IAA decay rate       &   \(\mu_r\)          & 5     &  \(min^{-1}\)       \\
Auxin biosynthesis rate    &  \(\alpha_a\)   & 0.5      &  \(\mu M min^{-1}\)     \\
AUX1 biosynthesis rate    & \(\alpha_u\)    & 5     &  \(\mu M min^{-1}\)     \\
AUX1 degradation rate        &   \(\mu_u\)          & 5      &   \(min^{-1}\)     \\
Rate of AUX1 localisation to membrane     &   \(\omega_u\)         & 0.5     &   \(\mu m \, min^{-1}\)     \\
Rate of AUX1 dissociation from membrane   & \(\delta_u\)    & 0.05        &  \(min^{-1}\)     \\
Maximum rate of PIN2 localisation to membrane    &   \(\omega_p\)      & 0.5     &  \(\mu m \, min^{-1}\) \\
Rate of PIN2 dissociation from membrane     & \(\delta_p\)   & 0.05        & \(min^{-1}\)  \\
Fraction of protonated auxin in cell   &  \(\kappa_a^{ef}\)     & 0.004      &  - \\
Fraction of protonated auxin in wall  &  \(\kappa_a^{in}\)     & 0.24       &  - \\
Effective PIN2-induced auxin efflux  &  \(\kappa_p^{ef}\)     & 4.67        &  - \\
Effective AUX1-induced auxin influx   &   \(\kappa_u^{in}\)   & 3.56        &  - \\
Auxin membrane permeability         &  \(\phi_a\)    & 0.55      &   \(\mu m \, min^{-1}\) \\
PIN2-induced auxin membrane permeability  &   \(\tilde{\phi}_p\)   & 0.27  & \(\mu m \, \mu M^{-1} min^{-1}\)  \\
AUX1-induced auxin membrane permeability  &  \(\tilde{\phi}_u\)    & 0.55  & \(\mu m \, \mu M^{-1} min^{-1}\)  \\
Rate of auxin diffusion in apoplast      & \(\phi_A\)      & 67    &  \(\mu m \, min^{-1}\) \\
Sensitivity of PIN2 localisation to auxin flux        & \(h\)    & 50   &  - \\
Auxin flux threshold    &  \(\theta\)  & 2  & -  \\

  \bottomrule
\end{tabular}
\end{table}

\begin{equation}
\label{delta_p_modified}
\delta_p^{modified} \bm{=} \delta_p \times \frac{P_a(t_f)}{P_a(t_i)},
\end{equation}

\noindent where \(t_i\) and \(t_f\) are the starting and ending times of the simulation.

\subsection{FER-ROPGEF-ROP6}

The FERONIA protein activation by apaoplastic auxin can be modeled using the Hill equation, i.e.,

\begin{equation}
\label{FER_Hill}
R_F \bm{=} \frac{A^2}{ k_F + A^2},
\end{equation}

\noindent where \(R_F\) represents the fraction of activated FERONIA proteins, and A represents the average auxin distribution in apoplast, averaging all four \(A_{ij}\) levels described in \cite{Mathematical_Modelling_of_Auxin_Transport_in_Plant_Tissues_Flux_Meets_Signalling_and_Growth}. Multiplying this ratio with the total FERONIA concentration, which is given in \autoref{Calculated_Values_for_the_Model}, will yield the active FERONIA concentration, \(F_a\). FERONIA protein is thought to play a role in the phosphorylation of ROPGEF, although this may be an indirect effect \cite{Functional_analysis_of_related_CrRLK1L_receptor_like_kinases_in_pollen_tube_reception}. Still, the previous phosphorylation model used for PIN2 phosphorylation by CPK29 (\ref{CPK29_PIN_Phosp}) is also applied to ROPGEF activation by the FERONIA protein. In this case, \(\nu_2\) is taken directly proportional to the activated FERONIA concentration, \(F_a\) shown as

\begin{equation}
\label{nu_2_ROPGEF}
\nu_2^R \bm{=}  8E11 \times F_a.
\end{equation}

The values for \(\nu_1^R\), \(j_1^R\) and \(j_2^R\) are given in \autoref{Fixed_Arbitrarily_Chosen_Values_for_Model_Parameters}. Applying a similar logic and utilizing the total ROPGEF concentration given in \autoref{Calculated_Values_for_the_Model}, the activated ROPGEF concentration, \(G_a\), is found. In this calculation, ROPGEF4 concentration is used specifically. Then, the Michaelis-Menten equation is used once more to model the ROP6 activation by ROPGEF as follows

\begin{equation}
\label{ROP6_production}
\upsilon_O \bm{=} \frac{\upsilon_{Pm} G_a(t)}{k_P + G_a(t)},
\end{equation}

\noindent where integration of \(\upsilon_O\) over time yields the active ROP6 concentration. Then, the variable \(\omega_p\)
which is related to the maximum rate of PIN2 localisation to membrane is modified for each cell as follows

\begin{equation}
\label{omega_p_modified}
\omega_p^{modified} \bm{=} \delta_p \times \frac{O_a(t_f)}{O_a(t_i)},
\end{equation}

\noindent where \(O_a\) represents the activated ROP6 concentration, as ROP6 activity is proposed to inhibit internalization of PIN2 through stabilization of actin filaments in roots \cite{A_ROP_GTPase-Dependent_Auxin_Signaling_Pathway_Regulates_the_Subcellular_Distribution_of_PIN2_in_Arabidopsis_Roots}.

\section{Simulation Results}

In this section, the cell wall and MCA2 channel blocks are inspected in detail. Moreover, some critical results of the simulation are provided. Some specific parameters and their values are given in \autoref{Calculated_Values_for_the_Model}, and fixed chosen parameter values of the model are shown in \autoref{Fixed_Arbitrarily_Chosen_Values_for_Model_Parameters}.

\begin{table}[!t]
  \caption{Calculated/Assigned Values for the Model}
  \centering
  \label{Calculated_Values_for_the_Model}

  \renewcommand{\arraystretch}{0.7}
  
  \begin{tabular}{@{}>{\raggedright\arraybackslash\itshape}p{4cm}>{\raggedright\arraybackslash\itshape}p{4cm}@{}}
    \toprule

  \textbf{Parameter}          & \textbf{Value} \\
 \midrule
\( A_M \)         & \(91106.18695\mathring{A}^2\) \\

\(\Delta x_d \)     &  7E-9 m\\

\(V_{h}\)  & 238.15 mV \\

\(\sigma_{h}\)    & 72.32 mmHg\\

\(\Delta E\)        &  150E-3 V\\

Total RBOHC  concentration &  1.420265781E-8 \, mol/L\\

Total CPK29  concentration &  1.724252492E-7 \, mol/L\\

Total FERONIA  concentration &  4.370431894E-7 \, mol/L\\

Total PIN2  concentration & 3.518272425E-8 \, mol/L\\

Total ROPGEF4  concentration &  4.634551495E-9 \, mol/L\\

Total ROP6  concentration &  5.830564784E-8 \, mol/L\\

F &  \(96485 \, C/mol\)\\

R &  \(8.314 \, J/mol·K\)\\

T &  \(298 \, K\)\\

V &  \(1E-5 \, L\)\\

\(G_{a}\) &  17E-12 \, S\\

\(l\) &  15.6871\\

  \bottomrule
\end{tabular}
\end{table}

\begin{table*}[!t]
  \caption{Chosen Values for Model Parameters}
  \centering
  \label{Fixed_Arbitrarily_Chosen_Values_for_Model_Parameters}

  \renewcommand{\arraystretch}{0.7}
  
  \begin{tabular}{@{}>{\itshape}p{2cm}>{\itshape}p{3cm}|>{\itshape}p{2cm}>{\itshape}p{3cm}|>{\itshape}p{2cm}>{\itshape}p{3cm}@{}}
    \toprule

 \midrule
\(k_F \) & 1E-6  mol/L      & \(N_1\)    & 100          & \(x_m\) & 1m \\
 \midrule
\(\mu_s\) & 0.01 Pa sec     & \(\mu_w\)    & 13E10 Pa sec    & \(\tau_w\) & 0.02 sec \\
 \midrule
\(\tau_s\) & 0.001 sec     & Y    & 0.5 Pa   & \(k_V\)   & 32.64  mV \\
 \midrule
 \(k_\sigma\)  & 16.13 mmHg    & \(c_c(t_i)\)    &  150 nM       & \(k_d\)  & \(0.01 Pa^{-1}\) \\
 \midrule
 \(A_C\)  & 1.9635E-17  \(m^2\)       & \(n_C\)    & 40  & \(k_C\)  & 1e-7 M \\
 \midrule
 \(\upsilon_{Hm}\) & 4E-5 \(mol L^{-1}sec^{-1}\)       & \(m_H\)    & 1E-9 mol/L      & \(\upsilon_{sm}\)  & 1E-5 \(mol L^{-1}sec^{-1}\) \\
 \midrule
\( m_s\) &  1E-4  mol/L  & \(n_a\)    & 40  & \( k_a\)  & 1.336E-8  mol/L \\
\midrule
z  & 2      & \(k_{eff_1}\)    & 0.3  & \(k_{eff_2}\)  & 0.16 \\
\midrule
\(k_K\) &  1E-23 mol/L  & \(\nu_1^P\)    & 5 \(mol L^{-1}sec^{-1}\)  & \(j_1^P\)  & 0.1 mol/L \\
\midrule
\(j_2^P\) &  5 mol/L  & \(\nu_1^R\)    & 5 \(mol L^{-1}sec^{-1}\)  & \(j_1^R\)  & 0.1 mol/L \\
\midrule
\(j_2^R\) &  5 mol/L  & \(\upsilon_{Pm}\)  & 0.1 \(mol L^{-1}sec^{-1}\) & \(k_P\)  & 0.05 mol/L \\
\midrule

  \bottomrule
\end{tabular}
\end{table*}

Simulations were performed for \(150\) seconds in MATLAB R2023b. Two different time steps were employed: \(0.0005\) \(s\) (\(2\) \(kHz\) sample rate) for real-time flow modeling (sound propagation and MCA2 response), and \(0.5\) \(s\) for internal cell processes. The \(0.5\) \(s\) step is based on the assumption that the MCA2 channels produce two responses per second.

\begin{figure}[!ht]
  \centering
  \subfloat[]{
    \includegraphics[width=0.45\linewidth]{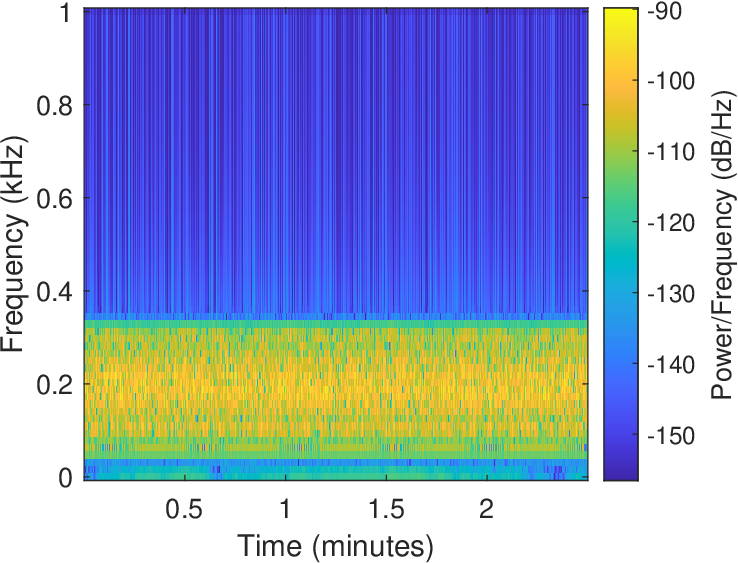}
    \label{fig:spect_in}}
  \hfill
  \subfloat[]{
    \includegraphics[width=0.45\linewidth]{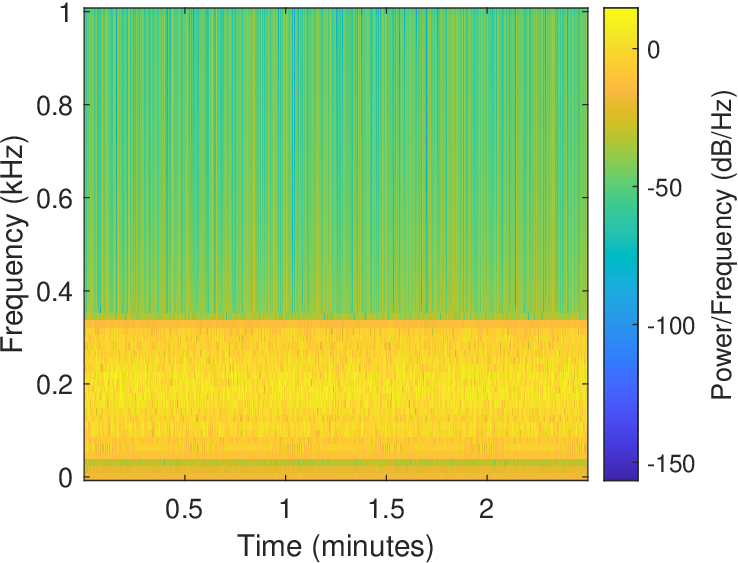}
    \label{fig:spect_out}}
  \caption{Spectrograms of \protect\subref{fig:spect_in} input and \protect\subref{fig:spect_out} output signals of the cell wall.}
  \label{fig_cell_wall_spect}
\end{figure}

\begin{figure}[!ht]
  \centering
  \includegraphics[width=0.6\linewidth]{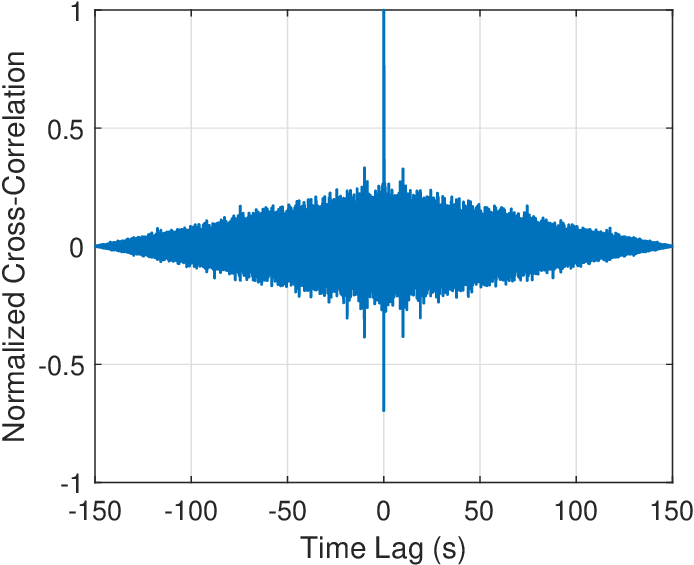}
  \caption{ Cross-Correlation of input and output signals of the Cell Wall.}
  \label{fig_Cell_wall_cross_corr}
\end{figure}

The cell wall takes stress (Pa) on the cell walls as its input and yields pressure (Pa) on one MCA2 channel. To understand the frequency characteristics of this block, a spectrogram of the input and output of the system is obtained as shown in \autoref{fig_cell_wall_spect}. It is observed that the cell wall enhances the frequency components in the 0-1000 Hz range. Moreover, from \autoref{fig_Cell_wall_cross_corr} one observes that the cell wall preserves much of the input timing and shape in the operating frequency since there is a high correlation with small time lags.

\begin{figure}[!ht]
  \centering
  \subfloat[]{
    \includegraphics[width=0.47\linewidth]{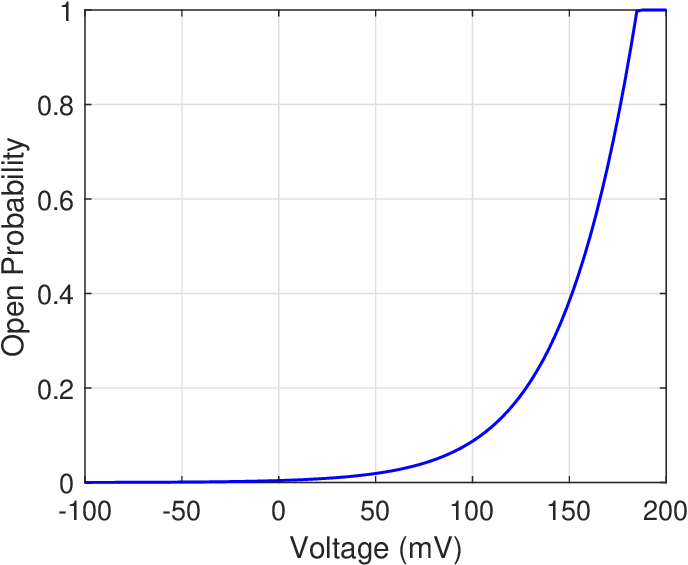}
    \label{fig:Open_Probability_vs_Voltage}}
  \hfill
  \subfloat[]{
    \includegraphics[width=0.47\linewidth]{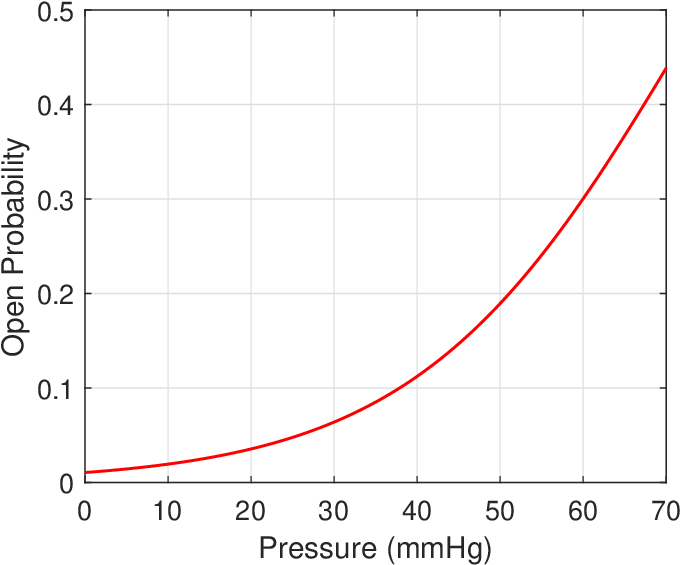}
    \label{fig:Open_Probability_vs_Pressure}}
  \caption{Open state probability characteristics of the MCA2 channel: 
  \protect\subref{fig:Open_Probability_vs_Voltage} variation with voltage at a pressure level of \(65\) \(mmHg\), and 
  \protect\subref{fig:Open_Probability_vs_Pressure} variation with pressure at \(150\) \(mV\).}
  \label{fig_Open_Probability_Plots}
\end{figure}

\begin{figure*}[!t]
  \centering
  \subfloat[]{
    \includegraphics[width=0.20\linewidth]{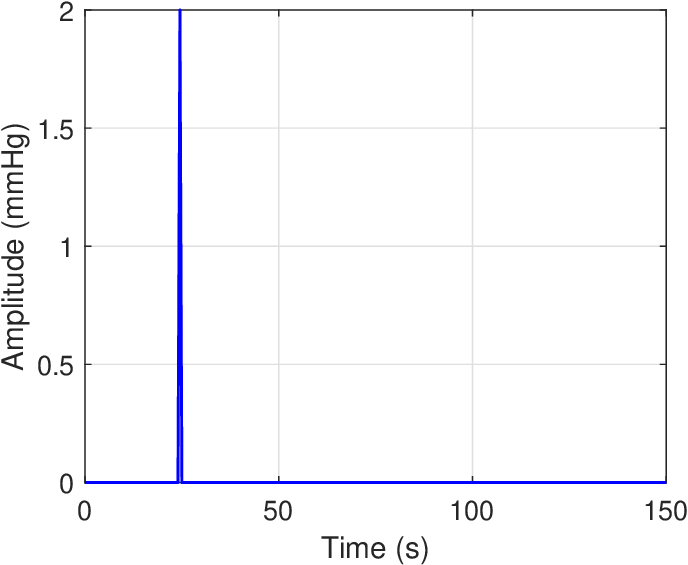}
    \label{fig:MCA2_impulse_in}}
  \hspace{0.015\linewidth}
  \subfloat[]{
    \includegraphics[width=0.20\linewidth]{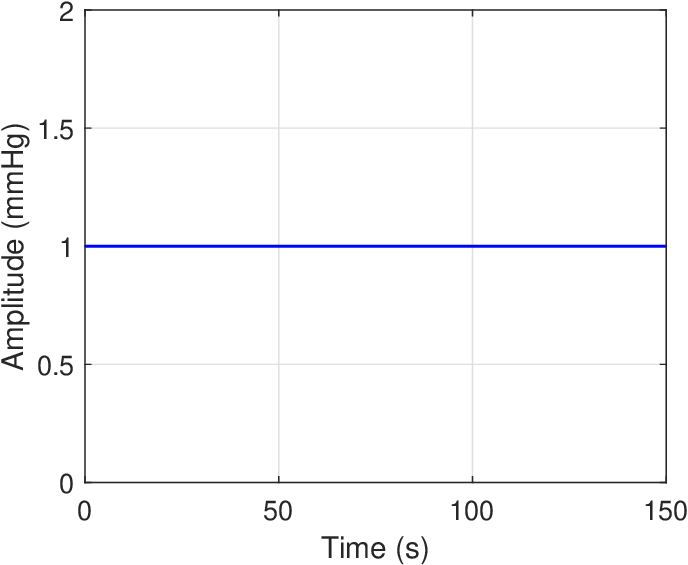}
    \label{fig:MCA2_step_in}}
  \hspace{0.015\linewidth}
  \subfloat[]{
    \includegraphics[width=0.20\linewidth]{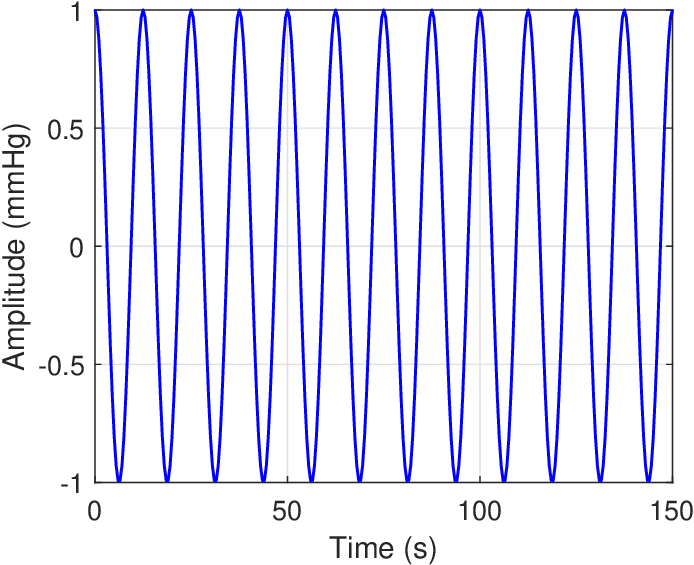}
    \label{fig:MCA2_cos_in}}
  \hspace{0.015\linewidth}
  \subfloat[]{
    \includegraphics[width=0.20\linewidth]{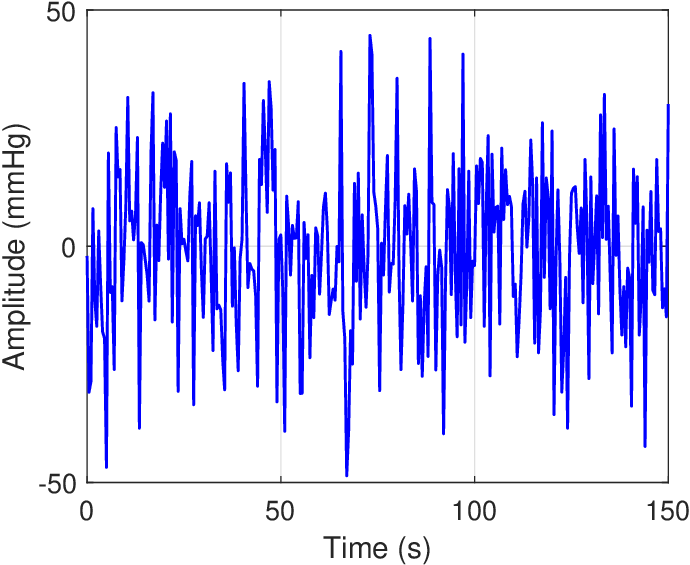}
    \label{fig:MCA2_general_in}}
  \\[1em]
  \subfloat[]{
    \includegraphics[width=0.20\linewidth]{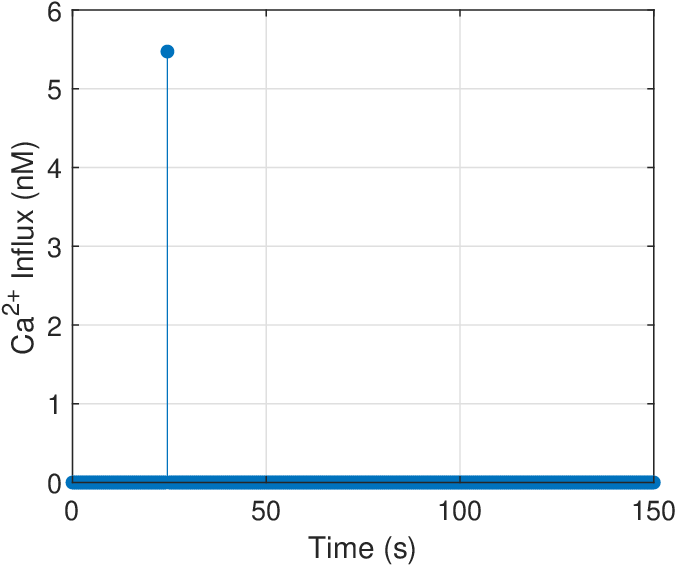}
    \label{fig:MCA2_impulse_out}}
  \hspace{0.015\linewidth}
  \subfloat[]{
    \includegraphics[width=0.20\linewidth]{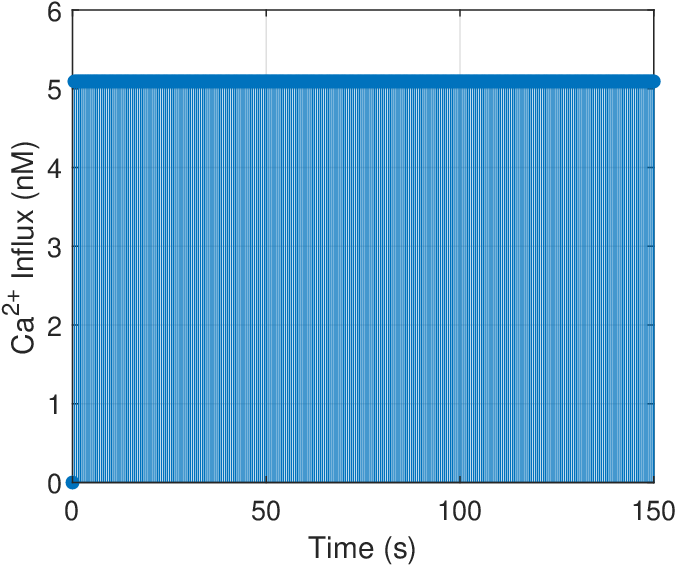}
    \label{fig:MCA2_step_out}}
  \hspace{0.015\linewidth}
  \subfloat[]{
    \includegraphics[width=0.20\linewidth]{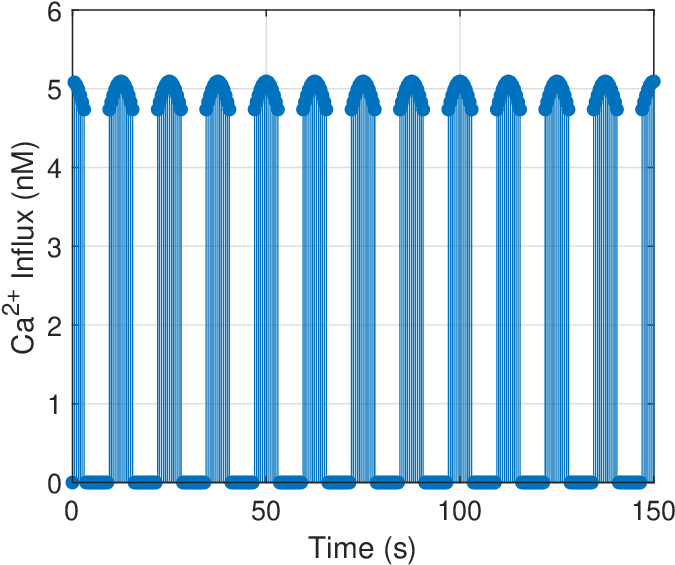}
    \label{fig:MCA2_cos_out}}
  \hspace{0.015\linewidth}
  \subfloat[]{
    \includegraphics[width=0.20\linewidth]{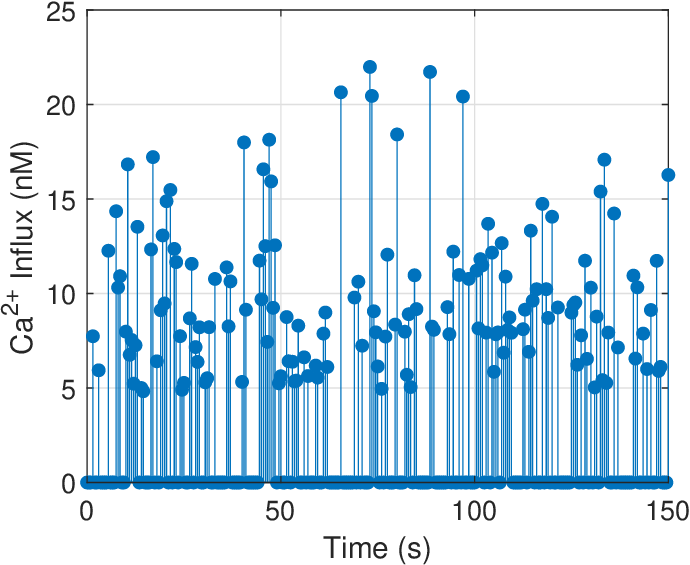}
    \label{fig:MCA2_general_out}}

  \caption{MCA2 channel responses to various input signals: 
    \protect\subref{fig:MCA2_impulse_in} impulse input and 
    \protect\subref{fig:MCA2_impulse_out} its output; 
    \protect\subref{fig:MCA2_step_in} step input and 
    \protect\subref{fig:MCA2_step_out} its output; 
    \protect\subref{fig:MCA2_cos_in} cosine input with frequency 0.08 Hz and 
    \protect\subref{fig:MCA2_cos_out} its output;
    \protect\subref{fig:MCA2_general_in} a possible input and
    \protect\subref{fig:MCA2_general_out} its output.}
  \label{MCA2_all_responses}
\end{figure*}

The MCA2 Channels block takes the pressure (\(mmHg\)) on an MCA2 channel as its input and produces the total \(Ca^{2+}\) influx concentration (\(nM\)) through the MCA2 channels. The lowpass filter in \autoref{MCA2_block_model} is a Butterworth filter with a cut-off frequency of \(250\) \(Hz\). The open probability of an MCA2 channel is plotted in \autoref{fig_Open_Probability_Plots}. Mathematically, after the filtering operation, the signal is downsampled to biological time steps. Moreover, the output of the MCA2 channel is characterized as a discrete-time signal. To understand the behavior of the MCA2 Channels block, different inputs are applied. Response to a shifted impulse is observed to result in only one spike, and the step response is observed to be another step function. The output to a sinusoidal input exhibits periodic oscillations with a rectified shape. Furthermore, a possible input-output signal pair is also calculated. These results can be observed in \autoref{MCA2_all_responses}. As it can be seen in \autoref{MCA2_all_responses}\subref{fig:MCA2_general_in}–\subref{fig:MCA2_general_out}, characteristics of \(Ca^{2+}\) influx through MCA2 channels are very similar to the current spikes given in \cite{MCAs_in_Arabidopsis_are_Ca2_permeable_mechanosensitive_channels_inherently_sensitive_to_membrane_tensions}.

\begin{figure}[!ht]
  \centering
  \includegraphics[width=0.6\linewidth]{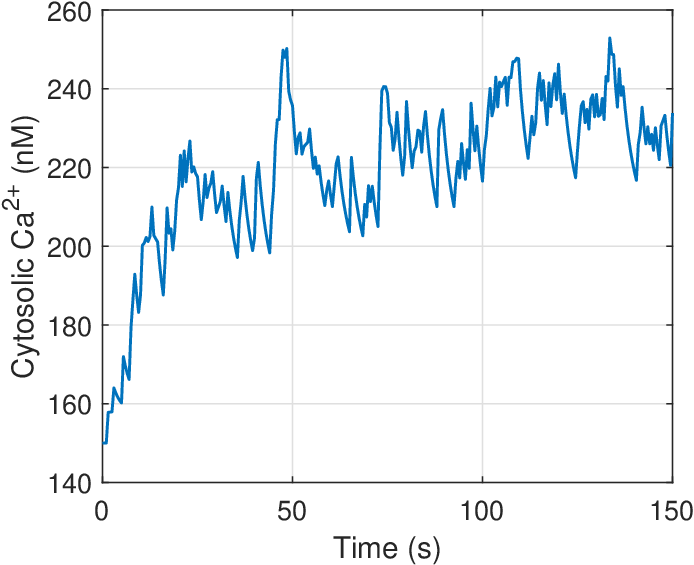}
  \caption{ Change of cytosolic \(Ca^{2+}\) concentration under sound stimuli through time.}
  \label{cytosolic_Ca_cont}
\end{figure}

In \autoref{cytosolic_Ca_cont}, the change in cytosolic calcium concentration is illustrated. Upon sound stimulation, a rapid increase in cytosolic \(Ca^{2+}\) levels is observed, facilitated by the \(Ca^{2+}\) - ROS Hub. Subsequently, the concentration stabilizes within the range of \(220\)–\(240\) \(nM\). These results are consistent with the findings reported in \cite{Root_phonotropism_Early_signalling_events_following_sound_perception_in_Arabidopsis_roots}.

\begin{figure}[!ht]
  \centering
  \includegraphics[width=0.6\linewidth]{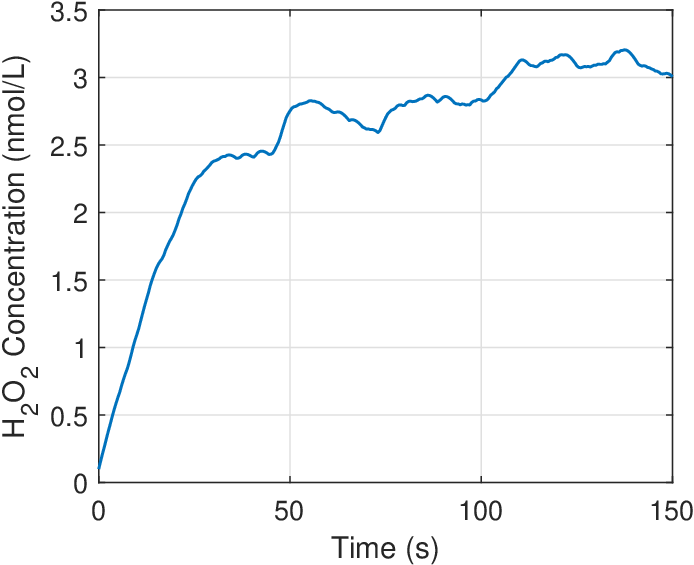}
  \caption{Change of apoplastic \(H_2O_2\) concentration under sound stimuli through time.}
  \label{H2O2_apop_cont}
\end{figure}

\autoref{H2O2_apop_cont} shows the change of apoplastic \(H_2O_2\) levels with time. The shape of this signal is very similar to the cytosolic \(Ca^{2+}\) concentration, since it also increases for some time and reaches a steady interval.


\begin{figure}[!ht]
  \centering
  \includegraphics[width=0.6\linewidth]{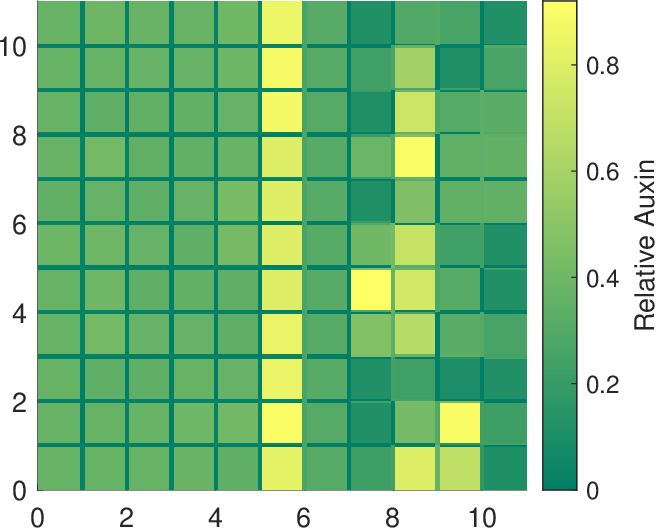}
  \caption{ Relative auxin levels of the 11x11 cell grid. Left side is assumed to be responding to sound stimuli while the right side is not.}
  \label{Grid_new}
\end{figure}

\autoref{Grid_new} shows an example 11x11 root cell grid where the middle column is assumed to be xylem where auxin flows from shoots to roots. Left-sided cells have modified \(\delta_p\) parameters as explained in (\ref{delta_p_modified}), representing the reaction to the sound stimuli. As can be seen, on the left side, the sound stimuli stops the flow of auxin while on the right side, some kind of auxin distribution occurs. With gravity, auxin will mostly gather on the edges of the roots on the right side, and those parts will grow faster resulting in the bending of roots towards the water flow.

\section{Communication-Theoretic Analysis of Plant Acoustic Sensing}

In this section, our objective is to quantify how variations in the characteristics of incoming acoustic stimuli influence the physiological responses they elicit in plants. Specifically, we investigate how changes in the mean frequency and mean amplitude of the sound wave affect plants' ability to elicit a response. Moreover, we examine how long it takes after the stimulus for the plant to generate a clear physiological response. To formalize these questions within a communication-theoretic framework, we performed a bit-error rate (BER) analysis.

From a communication-theoretic standpoint, the system can be treated as a digital link. A logical “1” represents the presence of water flow beneath the soil and triggers a sound wave generated by (\ref{Sound_underground_sum}); a logical “0” represents its absence, so no acoustic signal is emitted. The receiver declares bit 1 when a strong polar auxin distribution is present in the roots and declares bit 0 when no such distribution is observed. Simulations show that a visible polar distribution appears whenever the Activated PIN2 Ratio (APR), defined as the ratio of activated PIN2 concentration at the end of a bit interval to that at its start exceeds 5. Thus the decision rule can be written as

\begin{equation}
\label{decision_rule}
\hat{b} \bm{=} 
\begin{cases} 
 1, &  \text{if } APR > 5 \\
 0, &  \text{otherwise}
\end{cases}.
\end{equation}

\begin{figure}[!ht]
  \centering
  \subfloat[]{
    \includegraphics[width=0.45\linewidth]{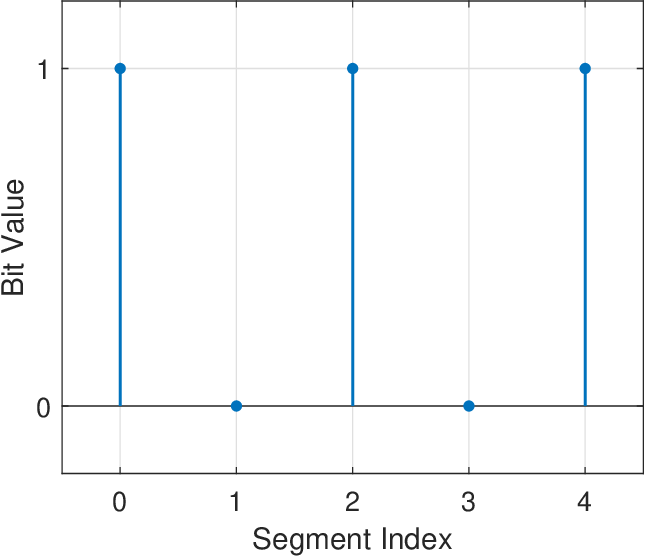}
    \label{fig:msg_tx}}
  \hspace{0.02\linewidth}
  \subfloat[]{
    \includegraphics[width=0.45\linewidth]{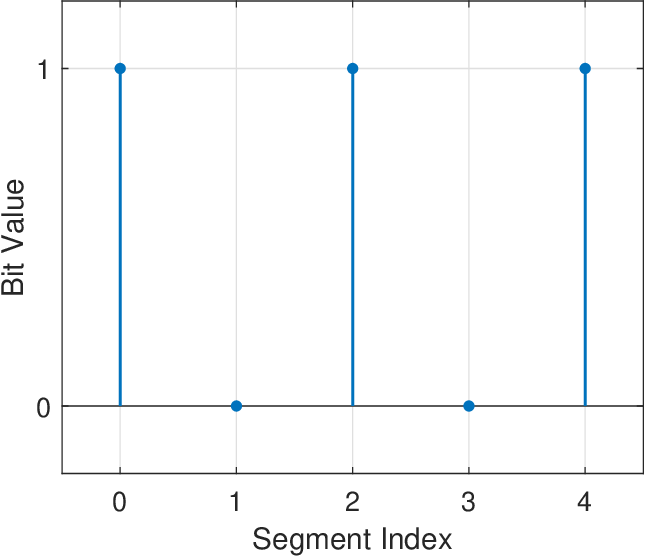}
    \label{fig:msg_rx}}
  \caption{\protect\subref{fig:msg_tx} Transmitted and \protect\subref{fig:msg_rx} received bits.}
  \label{comm_IO_bits}
\end{figure}

\begin{figure}[!ht]
  \centering
  \includegraphics[width=0.6\linewidth]{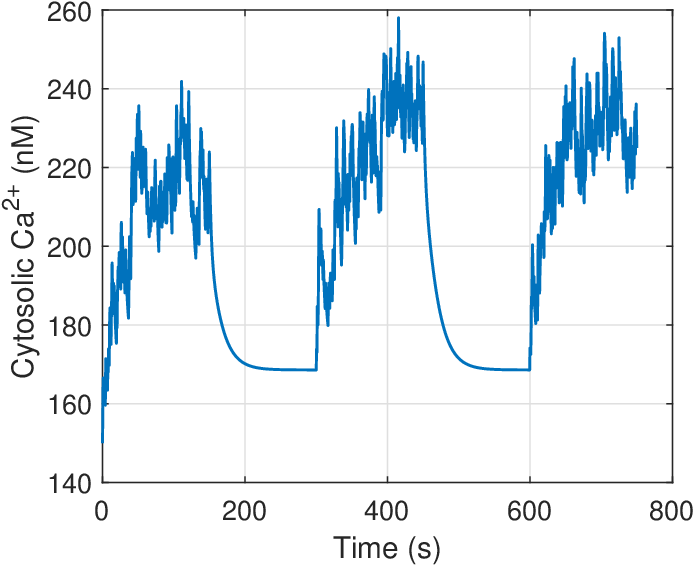}
  \caption{ \(Ca^{2+}\) concentration change due to sound waves.}
  \label{comm_cytos_Ca_cont}
\end{figure}

In \autoref{comm_IO_bits} an example input-output bit comparison is given. The corresponding cytosolic \(Ca^{2+}\) concentration plot is also provided in \autoref{comm_cytos_Ca_cont}. Here for one bit of information, the signal is sent for 150 seconds, and 5 bits are sent in total.

\begin{figure}[!ht]
  \centering

  \subfloat[]{
    \includegraphics[width=0.6\linewidth]{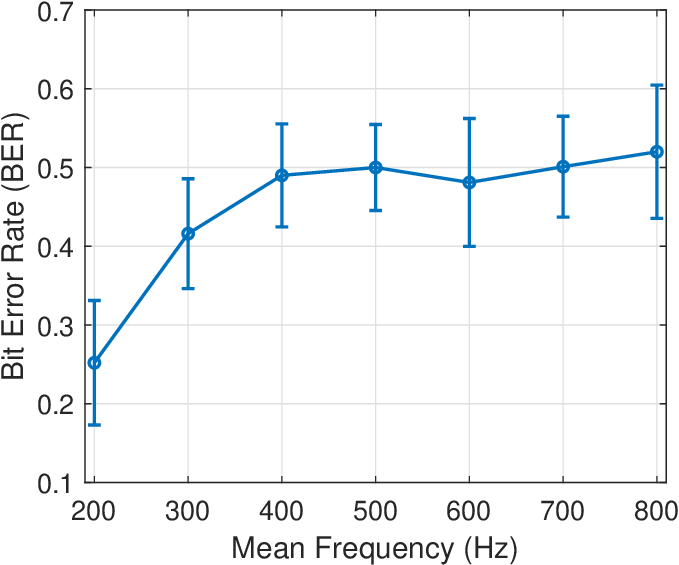}
    \label{fig:BER_vs_freq}}
  \\[0.8em]
  \subfloat[]{
    \includegraphics[width=0.6\linewidth]{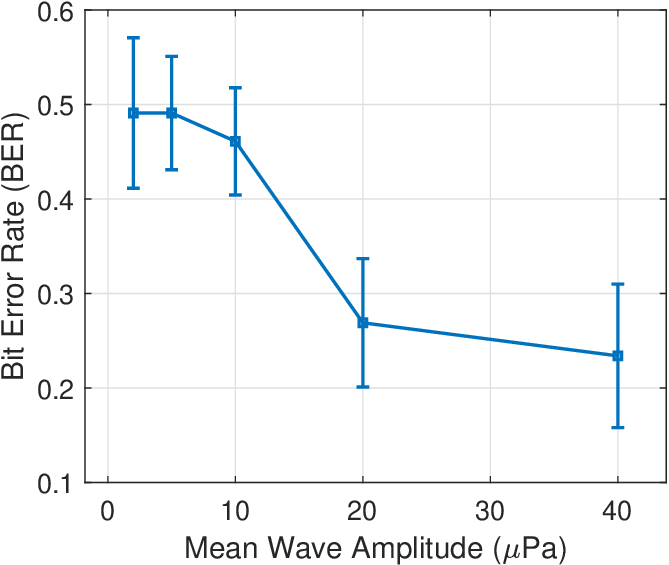}
    \label{fig:BER_vs_amp}}
  \\[0.8em]
  \subfloat[]{
    \includegraphics[width=0.6\linewidth]{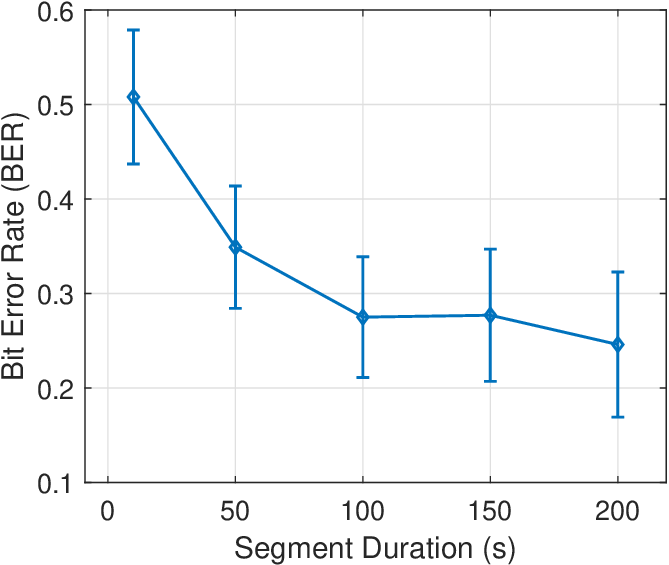}
    \label{fig:BER_vs_dur}}

  \caption{Bit Error Rate (BER) analysis under varying system parameters: 
  \protect\subref{fig:BER_vs_freq} mean frequency, 
  \protect\subref{fig:BER_vs_amp} mean amplitude, and 
  \protect\subref{fig:BER_vs_dur} plant decision time.}
  \label{BER_all}
\end{figure}

Using this concept, one can inspect the performance of the system with varying levels of sound frequency and amplitude. The performance of the system is evaluated as a function of the mean acoustic frequency. The mean of the Gaussian distributed carrier \(f_n\) is stepped through \([200, 300, 400, 500, 600, 700, 800]\,\text{Hz}\) where all other parameters are kept as their original values. For each mean value, 20 independent runs are carried out, each transmitting 50 bits and the bit-error rate (BER) is calculated. The results, shown in \autoref{BER_all}\subref{fig:BER_vs_freq}, reveal a pronounced increase in BER as the mean frequency increases from 200 Hz.

To investigate the effect of the sound amplitude \(A_s\) on the system performance, the mean of the Gaussian distributed amplitude level is iterated over \([2, 5, 1, 20, 40]\ \mu Pa\) where all other parameters are kept as their original values. Similarly, for each mean value, 20 independent runs are performed, each transmitting 50 bits and the bit-error rate (BER) is calculated. Results are shown in \autoref{BER_all}\subref{fig:BER_vs_amp}. It is observed that for small pressure levels, the plant is insensitive to sound waves.

Plant decision time describes how much time is needed for the plant to correctly decide on the input bit. Previously this was set to 150 seconds. To understand how fast plants can decide correctly whether there is a water flow, plant decision time is varied across \( [10, 50, 100, 150, 200] \) where all other parameters are kept as their original values. For each value, 20 independent runs are performed, each transmitting 50 bits and the bit-error rate (BER) is calculated. The resulting plot can be seen in\autoref{BER_all}\subref{fig:BER_vs_dur}. It becomes clear that the plant needs at least 100 seconds to decide whether there is water flow or not.

\section{Conclusion}
In this paper, a mathematical model describing how sound waves affect plants is proposed for the first time in the literature. Water flow under the soil is considered the source of the sound waves, the soil is the channel for the sound wave propagation, and the root cells are the receivers of the stimuli. How sound stimuli might trigger a response in plants is described by considering mechanosensitive channels, the Ca-ROS Hub proposal, and the role of auxin hormone in plants.

The analysis reveals that initially increased cytosolic calcium concentration can induce ROS production and create a positive feedback loop to increase cytosolic calcium concentration. Moreover, it is confirmed that PIN2 redistribution caused by sound stimuli may lead to the bending of roots.

Studying phytoacoustics requires an understanding of various aspects of plant biology, and there are many questions that need to be answered. More research is needed to address key areas, including:

\begin{itemize}
    \item How do plants extract different meanings from different sound waves?
    \item What is the evolutionary background of sound perception in plants?
    \item Can plants transmit meaningful messages using sound waves?
    \item Can plants communicate with each other in this manner?
    \item Do plants have any form of acoustic memory? If so, can this be transferred to subsequent generations?
    \item How exactly can farmers make use of sound wave stimuli?
\end{itemize}

Exploring this field may open new perspectives on plant behavior and shed light on how we can utilize sound waves to enhance plant growth and health.

\bibliographystyle{IEEEtran}
\bibliography{references.bib}

\vspace{11pt}

\begin{IEEEbiography}[{\includegraphics[width=1in,height=1.25in,clip,keepaspectratio]{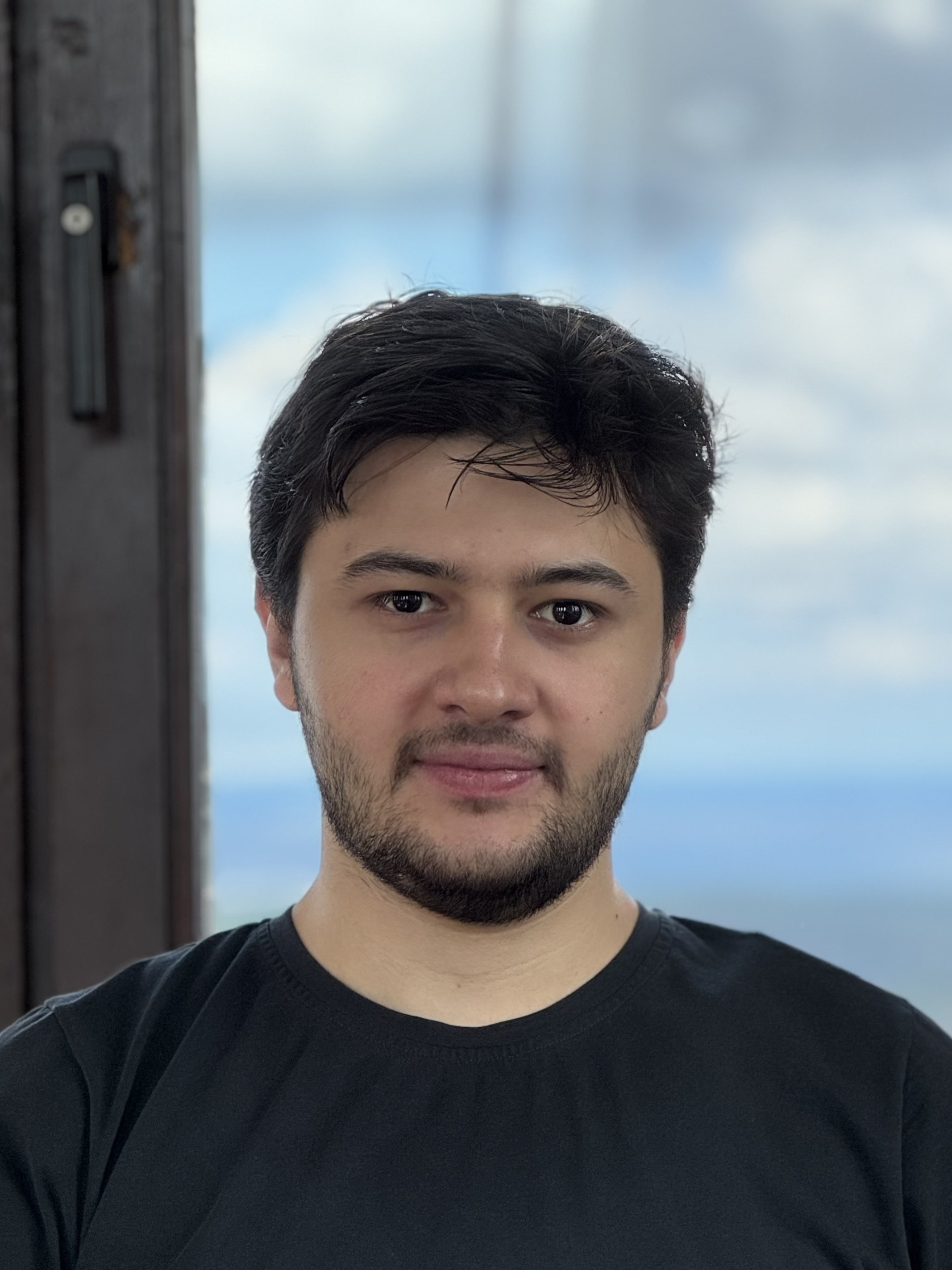}}]{Fatih Merdan}
completed his high school education at Kırıkkale Science High School. He received his B.Sc. degree in Electrical and Electronics Engineering from Middle East Technical University. He is currently pursuing his M.Sc. degree in Electrical and Electronics Engineering under the supervision of Prof. Akan at Koç University, Istanbul, Turkey.
\end{IEEEbiography}

\vspace{11pt}

\begin{IEEEbiography}[{\includegraphics[width=1in,height=1.25in,clip,keepaspectratio]{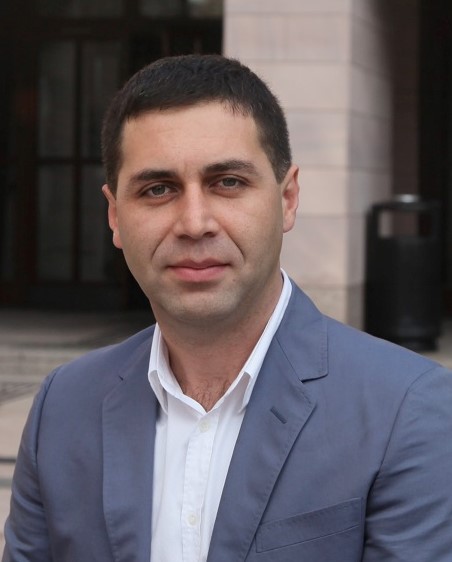}}]{Ozgur B. Akan}
\textbf{(Fellow, IEEE)} received the PhD
from the School of Electrical and Computer Engineering Georgia Institute of Technology Atlanta,
in 2004. He is currently the Head of Internet of
Everything (IoE) Group, with the Department of
Engineering, University of Cambridge, UK and the
Director of Centre for neXt-generation Communications (CXC), Koç University, Turkey. His research
interests include wireless, nano, and molecular communications and Internet of Everything.
\end{IEEEbiography}

\vfill

\end{document}